\documentclass[twocolumn]{autart}    % Enable this line and disable the 
\usepackage{mathtools}
\usepackage{amsmath} % assumes amsmath package installed
\usepackage{amssymb}
\usepackage{graphicx}          % Include this line if your 
\usepackage{upgreek}
\usepackage{xcolor}
\usepackage{comment}

\begin{document}

\begin{frontmatter}

\title{A Novel Four-DOF Lagrangian Approach to Attitude Tracking for Rigid Spacecraft}
%\title{In Catilinam IV\thanksref{footnoteinfo}} % Title, preferably not more 

\thanks[footnoteinfo]{Corresponding author: Yu Tang.}

\author[UNAM]{Eduardo Esp\' indola}\ead{eespindola@comunidad.unam.mx},    % Add the 
\author[UNAM]{Yu Tang}\ead{tang@unam.mx}             % e-mail address 
%\author[Baiae]{Publius Maro Vergilius}\ead{vergilius@culture.ir}  % (ead) as shown

\address[UNAM]{Faculty of Engineering, National Autonomous University of Mexico, Mexico City, 04510, MEXICO}  % Please supply                                              
%\address[Rome]{Senate House, Rome}             % full addresses
%\address[Baiae]{The White House, Baiae}        % here.

                             % keyword list or with the 
                                          % help of the Automatica 
                                          % keyword wizard

\begin{abstract}                          % Abstract of not more than 200 words.
    This paper presents a novel Lagrangian approach to attitude tracking for rigid spacecraft using  unit quaternions, where the motion equations of a spacecraft  are described by a four degrees of freedom Lagrangian dynamics subject to a holonomic constraint imposed by the norm of a unit quaternion. The basic energy-conservation property as well as  some additional useful properties of the Lagrangian dynamics are  explored,  enabling to develop quaternion-based attitude tracking controllers by taking full advantage of a broad  class of tracking control designs for mechanical systems based on energy-shaping methodology.  Global tracking of a desired attitude on the unit sphere is achieved by designing control laws that render the tracking error  on the four-dimensional Euclidean space  to converge to the origin. The topological constraints for globally exponentially  tracking  by a quaternion-based continuous controller and singularities in controller designs based on any three-parameter representation of the attitude  are then avoided.  Using this approach, a full-state feedback controller is first developed, and then several important issues, such as robustness to noise in quaternion measurements,  unknown on-orbit torque disturbances,  uncertainty in the inertial matrix, and  lack of angular-velocity measurements are addressed progressively,  by designing a hybrid state-feedback controller, an adaptive hybrid state-feedback controller, and an adaptive hybrid attitude-feedback controller. Global asymptotic stability is established for each controller. Simulations are included to illustrate the theoretical results. 

\end{abstract}

\end{frontmatter}

\section{Introduction}
Energy shaping based methodology was  a millstone along robot manipulator control designs \cite{takegaki1981new} and has been a basic building block for nonlinear control theory \cite{slotine1991applied,khalil2002nonlinear}. The main idea is to shape the potential energy of the underlined system by a proportional term in the controller such that the closed-loop system has a unique and isolated minimum at the origin where the tracking error is zero. The required damping to achieve the asymptotic stability is injected by a derivative action. The damping injection is also possible when only position is available for feedback through damping propagation.  A key to this methodology is to find an appropriate potential and kinetic energy for the closed-loop system. After that, a PD-like controller can be readily obtained by computing the gradient of the potential and  the kinetic energy.

Two approaches have been applied to control designs based on this methodology: Lagrangian and Euler-Newtonian approach.  In the Lagrangian approach, by exploring the physic properties of a Lagrangian system, i.e., the energy conservation characterized by the positiveness  of the mass/inertial matrix and the skew-symmetry of some  matrices  involved in the Lagrangian dynamics, it enables proposing energy-like Lyapunov function candidates.  The resulting controller consists of a PD action obtained  as the gradient of the potential energy and the kinetic energy of closed loop plus some form of feedforward compensation \cite{takegaki1981new,koditschek1984natural,spong1989robot,berghuis1994robust,kelly2006control,ortega2013passivity}.  Up to date a broad class of tracking control designs has been proposed in the literature and can be readily applied to problems of controller development  ranging from  basic full-state (position and velocity) feedback \cite{paden1988globally}, output (position) feedback \cite{berghuis1994robust}, adaptive  \cite{slotine1987adaptive,ortega1989adaptive,slotine1989composite,tang1994adaptive}, and robust compensations of uncertainties/external disturbances \cite{tang1998terminal,feng2002non,cruz2020continuous}, to decentralized control \cite{fu1992robust,tang2000decentralized,mei2012distributed} and coordination among a set of Lagrangian systems \cite{spong2007synchronization,chung2009application,mei2011distributed}. 

The Euler-Newtonian approach, on the other hand, has been an intensive research area for motion control of a rigid body. In particular, for spacecraft attitude control many important results have been presented. Commonly, the control law is designed on the special orthogonal group of dimension three $SO(3)$  using the rotation matrix \cite{bullo1999tracking,chaturvedi2011rigid}, on the unit sphere  $\mathcal S^3$ embedded in the space $\mathbb R^4$  using  unit  quaternions \cite{wen1991attitude,egeland1994passivity,lizarralde1996attitude,thienel2003coupled,luo2005inverse,tayebi2008unit},  or on the Euclidean space $\mathbb R^3$ using a 3-parameter representation   to parametrize the attitude \cite{tsiotras1998further,akella2001rigid,arjun2020uniform}. Due to the inherent  singularities, control designs based on a 3-parameter representation are limited to small rotations precluding global results, and may bring additional difficulties in planning the desired trajectory \cite{shuster1993survey}. 
%In quaternion based designs, the potential function depends   typically on the the vector part of the error quaternion by exploiting its norm constraint and the fact that two quaternions  correspond to the same physical rotation. This may, again,  lead to achieving only almost asymptotic or semi-global asymptotic stability, and also to  the  unwinding phenomenon. Controller designs based on a 3-parameter representation of  attitude with exponential stability have been reported in  \cite{arjun2020uniform}. However, like in the Lagrangian approach, attitude singularities appear necessarily in these controllers for using a 3-parameter representation of attitude. Indeed, the singularities may appear even in a singularity-free attitude representation such as rotation matrices \cite{lee2012exponential,lee2015global}. 
%To overcome theses difficulties, hybrid control strategy has been incorporated in the attitude tracking controller designs, and global stability has been achieved for configuration variables in the unit sphere  \cite{mayhew2011quaternion,schlanbusch2012hybrid,casau2019robust}, rotation matrix \cite{mayhew2013synergistic,lee2015global,berkane2017hybrid}, or quaternion logarithm \cite{espindola2021global}. 
The main difficulties  of using  rotation matrices  or unit quaternions  for attitude control designs, on the other hand,   are associated with the topological constraints  encountered  in the corresponding  group \cite{bhat2000topological}. Additionally, in quaternion-based designs,  it must  deal with care the ambiguity of unit quaternions in representing an attitude \cite{mayhew2011quaternion}.   These facts make the design of a potential function with a global minimum isolated critical point challenging.  Commonly used potential functions  are trace functions $\mathcal U(R)=tr\big(A(I-R)\big)$ in $SO(3)$ where $A\in \mathbb R^{3\times 3}$ is symmetric and  positive definite  \cite{koditschek1989application,mayhew2013synergistic,berkane2017hybrid}, or the height function  $\mathcal U(q)=1-|q|$ or its modified version in $\mathcal S^3$   \cite{wisniewski2003rotational,mayhew2011quaternion,casau2019robust}. The trace function creates,  besides the desired equilibrium corresponding to the desired attitude, unstable saddle equilibria at $\pi$ radiant rotations about the eigenvectors of the matrix $A$ \cite{mayhew2013synergistic}.  This gives at best the almost globally asymptotically convergent controller. This may have strong effects on the convergence of the tracking error, as the potential function approaches to zero when the attitude gets closer to one of the unstable equilibrium, causing a slow convergence rate \cite{lee2015global}.  The height function, on the other hand, aims at stabilizing  the attitude to one of two equilibria corresponding to the scalar part of the error quaternion equal to $\pm 1$ by a discontinuous control. However, this stability property is not robust to arbitrarily small measurement noise \cite{mayhew2011quaternion}. 

Recently, in \cite{mayhew2011synergistic,mayhew2013synergistic} and the references cited therein a family of smooth potential functions were constructed synergistically via angular warping on $SO(3)$, enabling  to design globally asymptotically  convergent controllers. The synergy property, which requires that for each undesired critical point of each potential function there exists a lower potential energy in the family, guarantees the robust global asymptotic tracking. Finding an explicit expression of the synergistic gap (the size of the hysteresis), however, is not straightforward.  More generic constructions of central synergistic  potential function \footnote{The centrality refers to the fact that all potential functions in the family share the same desired equilibrium \cite{mayhew2013synergistic}.} have  been advanced recently to facilitate this task  by applying a modified trace  function via introducing a perturbation to the tracking error when it reaches near to one of the undesired critical points while leaving the desired equilibrium unchanged \cite{berkane2017hybrid,berkane2018hybrid}. Similar ideas were applied for controller designs on $\mathcal S^3$ \cite{casau2019robust}. However, to achieve exponential tracking more restrictive conditions on the class of potential energy must be imposed. Also, gradient calculation of the potential energy is more involved and the parameters in the potential energy must be carefully selected \cite{berkane2018hybrid,casau2019robust}.

This paper presents a novel Lagrangian approach to attitude tracking for rigid spacecraft using  unit quaternions. In this approach,  the motion equations of a spacecraft are described by a four degrees of freedom (DOF) Lagrangian dynamics subject to a holonomic constraint imposed by the norm of a unit quaternion. The basic energy-conservation property as well as  some additional useful properties of the Lagrangian dynamics are  explored. Since the controller is developed on  the Euclidean space $\mathbb R^4$,  the resulting potential function is the same as used for control designs for mechanical systems on  Euclidean spaces, which has a unique isolated minimum in $\mathbb R^4$. Interestingly,  this potential function turns out  to be the height function in the unit quaternion group $\mathcal S^3$ used in \cite{mayhew2011quaternion} for robust attitude controller designs, and the switching is carried out when a simple switching condition is fulfilled.  This is in contrast to the potential functions commonly used for hybrid controller designs on $SO(3)$ \cite{mayhew2013synergistic,lee2015global,berkane2017hybrid} or $\mathcal S^3$ \cite{casau2019robust}, where a set of more elaborated potential functions was employed.

Compared to the approach presented in  \cite{fjellstad1994position,wu2011quaternion}, where the resulting Lagrangian dynamics verifies only part of  the properties of a Lagrangian system leading to local asymptotic stability of the closed loop due to the restrictions to a subspace of the configuration in the control design,  the 4-DOF Lagrangian dynamics presented  in this paper possesses the basic energy-conservation property as well as some  additional properties, enabling to design globally exponentially stable controllers in $\mathbb R^4$ for attitude tracking.  In contrast to the 3-DOF Lagrangian approach to attitude control using either the vector part of a quaternion \cite{caccavale1999output,costic2001quaternion} or a 3-parameter parametrization of the attitude \cite{tomei1992nonlinear,fossen1991adaptive,slotine1990hamiltonian,wong2001adaptive,chung2009application}, the proposed approach leverages the full quaternion to describe the attitude dynamics and therefore allows to design globally tracking controllers. Also, the topological constraints  encountered in the quaternion group  and  the singularities in any 3-parameter representation of the attitude are avoided.  

%Also, by exploring the basic energy conservation property and the additional properties of the Lagrangian dynamics, the proposed approach allows  for developing attitude tracking controllers using the quaternion parametrization by taking full advantage of a broad  class of tracking controller designs for robot manipulators available in the literature.

%Compared the synergistic potential functions in \cite\cite{mayhew2013synergistic,lee2015global,berkane2017hybrid,casau2019robust}, the potential function used here is simpler, and was commonly applied in attitude control \cite{mayhew2011quaternion,wisniewski2003rotational}. It allows to derive a simple switching condition for hybrid controller design.

%{\color{red}   % \rule{\linewidth}{0.5mm} } Assumptions  are three general properties
%for Euler–Lagrange systems. Examples include robot manipulators in joint space with unknown but constant masses, inertia and distances of the centers of mass of the links (\cite{kelly2006control}; \cite{spong1989robot}), attitude dynamics of rigid bodies with unknown but constant inertia (\cite{slotine1991applied}), and autonomous vehicles, to name a few.
%}

\subsection{Related works}
Compared to the amount of results for control designs for general mechanical  systems in Euclidean spaces, there is   few work based on the Lagrangian approach  addressing the problems of attitude control of a rigid body moving in a three-dimensional space, where the attitude configuration variables evolve in the rotation group of $SO(3)$ or the unit sphere $\mathcal S^3$ embedded in $\mathbb R^4$. In early works  the Lagrangian dynamics that describes the attitude employs  typically a 3-DOF formulation, obtained either by taking only the vector part of the quaternion into the configuration variable  \cite{caccavale1999output,costic2001quaternion}, or adopting a 3-parameter representation of the attitude, like Euler angles \cite{tomei1992nonlinear,fossen1991adaptive}, Rodriguez parameters (RP) \cite{slotine1990hamiltonian} or Modified Rodriguez Parameters (MRP) \cite{wong2001adaptive,chung2009application}. The energy-conservation property is explored in the underlined  Lagrangian dynamics for control  designs. An important exception was reported in  \cite{fjellstad1994position,wu2011quaternion}, where a  4-DOF Lagrangian dynamics was derivated based on the kinematics and dynamics. The 4-DOF Lagrangian dynamics verifies, nonetheless,  only part of  the properties of a Lagrangian system. This allows to conclude only local asymptotic stability of the closed loop due to restrictions to a subspace of the configuration in the controller design. In a recent work \cite{culbertson2021decentralized} a 9-DOF Lagrangian dynamics in $SO(3)$ was used to address attitude manipulation in collaborative robots.

%As commented above, the main difficulties  of using unit quaternions  for attitude controller development  are associated with the topological constraints  encountered  in the quaternion group and dealing with the unwinding phenomenon as the consequence of the ambiguity of the unit quaternion in representing an attitude. Any  3-parameter representation of the attitude, on the other hand, faces the singularity problem and only attitude trajectory with small rotation can be tracked.  To the authors' best knowledge, the early efforts in describing the attitude by a 4-DOF Lagrangian dynamics using the unit quaternion was reported in \cite{fjellstad1994position}. The Lagrangian dynamics obtained in this reference, however, verifies only part of  the properties of a Lagrangian system, which restricts the controller design to a subspace of the configuration, and only asymptotic stability of the closed loop was achieved.   

Recent advances in attitude control designs based on a 4-DOF Lagrangian dynamics have been reported in \cite{udwadia2012unified,rodriguez2020lagrange}. The Lagrangian dynamics is derivated by appending the holonomic constraints of a unit quaternion  to the Lagrangian using Lagrange multipliers leading to Euler-Lagrange equations that include Lagrange multiplies, which  are considered in conjunction with the algebraic constraint equations \cite{udwadia2010alternative,sherif2015rotational,lee2017global,rodriguez2020lagrange}. This approach obtains a set of configurations that satisfy the holonomic constraints  embedded in $\mathbb R^{4}$. The standard variational methods is applied while the variations are constrained to respect the geometry of the configuration manifold. However, control designs based on energy-shaping methodology are not addressed in these works, since the energy-conservation property is not explored in this approach.

\subsection{Contributions}
To address the aforementioned challenges in quaternion-based attitude control designs, this paper proposes to design attitude controllers on the Euclidean space $\mathbb R^4$. Towards this aim, 

% The contributions of this paper are summarized as follows:
\begin{enumerate}
    \item  A novel 4-DOF Lagrangian dynamics using  unit quaternions subject to the norm constraint for attitude control is derivated, the basic energy-conservation property of the Lagrangian dynamics as well as the additional useful properties are established.  Similar to the approach followed in \cite{caccavale1999output,fjellstad1994position,costic2001quaternion,wu2011quaternion}, the Lagrangian dynamics is derivated from the kinematic and dynamic equation of the rigid body. From the control design perspective, this approach is more suitable for derivation of control-orientated Lagrangian dynamics as compared to the approach of modeling the attitude using a Lagrangian function in conjunction with Lagrange  multipliers \cite{udwadia2010alternative,udwadia2010equations,rodriguez2020lagrange}, since it allows to explore the basic energy-conservation property and the additional properties of the Lagrangian dynamics, enabling to take full advantage of well developed control designs based on the energy-shaping  methodology for a broad class of mechanical systems. To the best of the authors' knowledge these properties are not revealed before in the literature for the attitude dynamics.

\item  The attitude tracking problem is then formulated on the Euclidean space $\mathbb R^4$ based on the 4-DOF Lagrangian system and  solved using  the energy-shaping methodology \cite{slotine1991applied}. 
Given a desired attitude trajectory in terms of a quaternion,  the spacecraft quaternion described by the 4-DOF  Lagrangian dynamics evolves on the unit sphere $\mathcal S^3$ embedded in the space $\mathbb{R}^4$.  The controller is designed to render the attitude tracking error, calculated as the Euclidean distance between the desired quaternion and the spacecraft quaternion, to converge to the origin of $\mathbb R^4$. The error quaternion, calculated as the quaternion product between the conjugate of the desired quaternion and the spacecraft quaternion, converges to the unit quaternion as the attitude tracking error converges to the origin of the space $\mathbb{R}^4$. 

\item The basic full state-feedback controller design is extended to address several important practical issues, such as robustness to noise in quaternion measurements,  unknown on-orbit torque disturbances, uncertainty in the inertial matrix, and  lack of angular-velocity measurements, by designing  progressively  a hybrid state-feedback controller, an adaptive hybrid state-feedback controller, and an adaptive hybrid attitude-feedback controller. Global asymptotic stability is established for each controller.   Since the controller is developed on the space $\mathbb R^4$,  the resulting  potential functions  have  a unique isolated minimum in the corresponding Euclidean space. The term in the  potential functions for the attitude error is the same as that  used in \cite{mayhew2011synergistic} for attitude control designs in $\mathcal S^3$, and the switching is carried out when a simple switching condition is fulfilled.  This is in contrast to the potential functions commonly used for hybrid controller designs in $SO(3)$ \cite{mayhew2013synergistic,lee2015global,berkane2017hybrid} or $\mathcal S^3$ \cite{casau2019robust}, where a set of more elaborated potential functions were employed.

%The basic full state-feedback controller design is extended to address several important problems, such as robustness to noises in quaternion measurements by a hybrid control, lack of angular velocity measurements by a nonlinear filter, and uncertainty in the inertial matrix by an adaptive control. Since the controller is developed in the space $\mathbb R^4$,  the resulting potential function is the same as used for controller designs for robot manipulators, which has a unique isolated minimum in $\mathbb R^4$. This potential corresponds to the potential function used in \cite{mayhew2011synergistic} for attitude controller designs in $\mathcal S^3$, and the switching is carried out in the group when a simple switching condition is fulfilled.  This is in contrast to the potential functions commonly used for hybrid controller designs in $SO(3)$ \cite{mayhew2013synergistic,lee2015global, berkane2017hybrid} or $\mathcal S^3$ \cite{casau2019robust}, where a set of potential functions were employed whose design is more complicated. 

\end{enumerate}

\subsection{Organization}
The rest of the paper is organized as follows. In Section \ref{Sec:Prel},  the rotation kinematics and dynamics of a rigid spacecraft  and some useful results for the Lagrangian dynamics derivation  are presented. In Section \ref{Sec:SF} the novel 4-DOF Lagrangian dynamics for the attitude of a rigid spacecraft using  unit quaternions is derivated  from the kinematic and dynamic equation of the spacecraft, the basic energy-conservation property of the Lagrangian dynamics (Lemma \ref{lem2}) as well as the additional properties (Lemma \ref{lem3}) are established, which allow to take full advantage of the well-developed control design methodologies based on energy-shaping for mechanical systems in problems related with the attitude control of a rigid body.  Section \ref{Sec:SF} devotes to  design a full state-feedback controller (Theorem \ref{thm1}). Extensions are presented  in this section  to address the issues  of robustness to noises in the quaternion measurements (Theorem \ref{clr1}), the presence of on-orbit torque disturbances and  uncertainty in the inertial matrix (Theorem \ref{thm2}), and lack of  angular-velocity measurements (Theorem \ref{thm3}).  In Section \ref{Sec:Sim}, numerical simulations are included to illustrate the performance of the proposed hybrid controllers subject to noisy attitude measurements, unknown inertial matrix and torque disturbances, and lack of angular-velocity measurements. Section \ref{Sec:Conc} draws conclusions. Proofs of the properties of the proposed Lagrangian dynamics are given in Appendixes. 

\subsection{Notations}
In this paper, the $l_2$-norm for a vector $u\in\mathbb{R}^n$ is denoted by $\|u\|$. Its induced matrix norm (induced spectral norm) for a matrix $A\in \mathbb{R}^{n\times n}$ ($A\in \mathbb{R}^{m\times n}$) is denoted by  $\| A\|$. $\lambda_{\max}(A)$ and $\lambda_{\min}(A)$ denote respectively the maximum and minimum eigenvalue of a positive definite matrix $A$. $I_n$ is the identity matrix of $n\times n$, and $0_{n\times m}$ is a matrix of $n\times m$ with zero in all its entries.  The map $S(\cdot): \mathbb{R}^{3} \to \mathfrak{so}(3)$ represents the cross-product operator $S(u)v=u\times v, \ \forall u,v\in \mathbb R^3$, where  $\mathfrak{so}(3) :=\lbrace A\in\mathbb{R}^{3\times 3  }| A = -A^{T} \rbrace$ is the set of skew-symmetric matrices of $3\times 3$.

\section{Rotational Dynamics of a Rigid Body} \label{Sec:Prel}
This paper uses unit quaternions $q$ to represent the attitude of the body frame $\mathbf{B}$ fixed to the mass center of the spacecraft respect to the inertial reference frame $\mathbf{I}$ fixed to the center of the Earth, where
\begin{equation*}
q=\left[ q_0 , \ q^{T}_{v} \right]^{T} \in \mathcal{S}^{3}, \; q_{0} \in \mathbb{R}, \; q_{v} \in \mathbb{R}^{3},
\end{equation*}
and $\mathcal{S}^{3}=\lbrace x \in \mathbb{R}^{4} | x^{T}x=1 \rbrace$ is the three-dimension unit sphere embedded in $\mathbb{R}^{4}$. The unit sphere doubly  covers the rotation  group, i.e.,   $q$ and $-q$ represent the same physical orientation. This can be noticed from  the Rodriguez formula  $R(q)=I_3 + 2q_{0}S(q_{v}) + 2S^{2}(q_{v})$ that $R(q)=R(-q)$. 

The motion equations of a spacecraft can be described by the Euler-Newton equation
\begin{align}
M\dot{\omega} &= S(M\omega )\omega + \tau  \label{eq:Dynamics}
\end{align}
and the kinematic equation
\begin{align}
\dot{q}&=\frac{1}{2}J(q)\omega , \label{eq:KinematicsQ}
\end{align}
where $\omega \in \mathbb{R}^{3}$ is the angular velocity, $\tau \in \mathbb{R}^{3}$ the applied torque, and $M\in\mathbb{R}^{3\times 3}$, $M =M^{T} >0$ denotes a constant, symmetric and positive definite inertial matrix
\begin{equation}\label{eq:InertialM}
    M = \left[ 
    \begin{array}{ccc}
         m_{11}& m_{12}&m_{13}  \\
         m_{12}&m_{22}&m_{23} \\
         m_{13}&m_{23}&m_{33}
    \end{array}
    \right],
\end{equation}
all expressed in the body frame. Matrix $J(q)\in\mathbb{R}^{4\times 3}$ in the kinematics \eqref{eq:KinematicsQ} is 
\begin{equation} \label{eq:JH}
J(q)=
\left[ 
\begin{array}{c}
-q^{T}_{v}  \\
q_{0} I_3 + S(q_{v})
\end{array}
\right].
\end{equation}
Some useful properties of matrix $J(\cdot)$  \cite{markley2014fundamentals} are listed below.

\emph{\textbf{Properties of matrix $J(\cdot)$:}}\label{PropJx}
For all $x, y\in \mathbb{R}^{4}$, the following properties hold:
\begin{enumerate}
\item  $J(\alpha x + \beta y)=\alpha J(x) + \beta J(y), \; \forall \alpha, \beta \in \mathbb{R} $. 
%\item $J^T(x)J(y)=x^Ty I_3 +\Big( x_vy_v^T-y_vx_v^T+x_0S(y_v)-y_0S(x_v)\Big)$. \ \     $J^{T}(x)J(x)=||x||^{2} I_{3}$, \label{pA1}
\item $J^{T}(x)y = -J^{T}(y)x$. \label{pA2}
\item $J^{T}(x)y = 0_{3\times 1} \iff y = kx,\; \forall k\in\mathbb{R}$. \label{pA3}
\item $J^{T}(x)J(x)=||x||^{2} I_{3}$. \label{pA1}
\item $\|J(x)\| =\| x\|$.
%\item $\|J^T(x)J(y)\|\leq \|J(x)\|\|J(y)\| =\|x\| \|y\|$.
\item $\frac{d}{dt}\left( J(x) \right) = J(\dot{x})$.
\end{enumerate}

Define the matrix $Q(x)\in \mathbb{R}^{4\times 4}$ as
\begin{equation}\label{eq:MatQ}
Q(x)\vcentcolon =
\left[ 
\begin{array}{cc}
x_0 & -x^{T}_{v}  \\
x_{v} & x_{0} I_3 + S(x_{v})
\end{array}
\right] =\left[ x \; J(x) \right]
\end{equation}
for any $x=\left[ x_{0},\right.$ $\left. x^{T}_{v} \right]^{T}$ $\in$ $\mathbb{R}^{4}$, with $x_{0} \in \mathbb{R}$, $x_{v} \in \mathbb{R}^{3}$.

\begin{lem}{\emph{\textbf{(Properties of matrix $Q(\cdot)$):}}}\label{lem1}
Matrix $Q(\cdot)\in \mathbb{R}^{4\times 4}$ defined in  \eqref{eq:MatQ} verifies the following properties $\forall x,y\in\mathbb{R}^{4}$
\begin{enumerate}
\item $Q(x)\in SO(4), \; \forall x\in \mathcal{S}^{3}$. \label{pQ1}
\item $Q(y)Q^{T}(x) = J(y)J^{T}(x) + yx^{T}$. \label{pQ2}
\item $u^{T}Q(y)Q^{T}(x)u = 0 \iff y^{T}x = 0$, $\forall u\in\mathbb{R}^{4}$. \label{pQ3}
\item $Q(y)Q^{T}(x) = -Q(x)Q^{T}(y) \iff y^{T}x = 0$. \label{pQ4}
\item $Q(\alpha x + \beta y)=\alpha Q(x) + \beta Q(y), \; \forall\alpha, \beta \in \mathbb{R} $.\label{pQ5}
\item $\frac{d}{dt}\left( Q(x) \right) = Q(\dot{x})$. \label{pQ6}
%\item $J(x)AJ^{T}(x) = Q(x)\bar{A}Q^{T}(x), \;\forall A\in\mathbb{R}^{3\times 3},\; \bar{A} =\left[ \begin{array}{cc}
%0 & 0_{1\times 3} \\
%0_{3\times 1} & A
%\end{array}\right] \in\mathbb{R}^{4\times 4}.$\label{pQ7}
\end{enumerate}
\end{lem}
\begin{pf}
See Appendix  \ref{AppQ}.
\begin{flushright}
$\square$
\end{flushright}
\end{pf}

\section{Derivation of the 4-DOF Lagrangian Dynamics}
By the kinematics \eqref{eq:KinematicsQ} and   Property \ref{pA1} of matrix $J(q)$, the angular velocity can be obtained as
\begin{equation}\label{eq:AngVel}
\omega = 2J^{T}(q)\dot{q}.
\end{equation}
Taking the time derivative of the kinematics \eqref{eq:KinematicsQ}, by \eqref{eq:Dynamics} and \eqref{eq:AngVel},  gives
\begin{align}\label{eq:qdd}
\ddot{q}=&\frac{1}{2}J(q)\dot{\omega} + \frac{1}{2}J(\dot{q})\omega \nonumber\\ 
 =& J(q)M^{-1}S(M\omega )J^{T}(q)\dot{q} + J(\dot{q})J^{T}(q)\dot{q}  \nonumber \\
 &+\frac{1}{2}J(q)M^{-1}\tau.
\end{align}
Let $m_{0}>0$ be a positive constant. Define 
\begin{equation}\label{eq:M0}
M_{0} \vcentcolon = \left[ 
\begin{array}{cc}
m_{0} & 0_{1\times 3 } \\
0_{3\times 1} & M
\end{array}\right],
\end{equation}
which is symmetric and positive definite, and let
\begin{equation}\label{eq:Dq2}
D(q)  := Q(q)M_{0}Q^{T}(q).
\end{equation}

%From the definition of $Q(q)$, this matrix can be expressed as
%\begin{equation}\label{eq:Dq}
%D(q) \vcentcolon = J(q)MJ^{T}(q) + m_{0}qq^{T} \in\mathbb{R}^{4\times 4}.
%\end{equation}
Premultiplying the matrix $D(q)$ in both sides of \eqref{eq:qdd}, by  Property \ref{pA3} ( $q^{T}\dot{q} = 0$) of the matrix $J(q)$ and after some algebraic manipulations, the attitude dynamics of the spacecraft can be described by the following 4-DOF Lagrangian dynamics
\begin{equation}\label{eq:EL-model} 
D(q)\ddot{q} + C(q,\dot{q})\dot{q} =  \bar{\tau},
\end{equation}    
with the positive definite inertial-like matrix $D(q)$ in \eqref{eq:Dq2} and Coriolis-centrifugal-like matrix defined as
\begin{equation}
C(q,\dot{q}) := -J(q)S(M\omega )J^{T}(q) - D(q)Q(\dot{q})Q^{T}(q). \label{eq:Cqqp}
\end{equation}
 The generalized torque in the Lagrangian dynamics is 
\begin{equation}
\bar{\tau} = \frac{1}{2}J(q)\tau. \label{eq:TauBar}
\end{equation}
Notice that the applied control torque $\tau$ to the spacecraft is obtained  by premultiplying in both sides the matrix $J^{T}(q)$ in \eqref{eq:TauBar}
\begin{equation}
\tau = 2J^{T}(q)\bar{\tau}.  \label{eq:Tau}
\end{equation}

\begin{lem}\label{lem2}
\emph{\textbf{(Basic properties of the Lagrangian dynamics \eqref{eq:EL-model}):}}
The spacecraft motion equations described by the Lagrangian dynamics \eqref{eq:EL-model} with $D(q)$  in \eqref{eq:Dq2} and  $C(q,\dot{q})$  \eqref{eq:Cqqp} have the following properties:
\begin{enumerate}
    \item Matrix $D(q)$ is symmetric and positive definite: 
   \begin{equation} \label{pLg1} 
   \underbar{m}I_{4}\leq D(q)\leq \bar{m} I_{4}, \ \ \forall q\in\mathcal{S}^{3},
\end{equation}    
where
\begin{align*}
\bar{m} &\vcentcolon = \max \left\{ m_{0} , \lambda_{\max}(M) \right\} \\
\underbar m &\vcentcolon = \min \left\{ m_{0},\lambda_{\min}(M) \right\}.
\end{align*}

    \item Matrix $\dot{D}(q) - 2C(q,\dot{q})$ is skew-symmetric:
\begin{equation}
x^{T} \left( \dot{D}(q) - 2C(q,\dot{q}) \right) x = 0, \ \ \forall q\in\mathcal{S}^{3} \ \mathit{and } \  x\in {R}^{4}. \label{pLg2}
\end{equation}    
\end{enumerate}
\end{lem}
\begin{pf}
See Appendix \ref{PfLemma2}.
\begin{flushright}
$\square$
\end{flushright}
\end{pf}

These basic properties are a reflection of the energy-conservation property of the rigid body in the Lagrangian dynamics, and  are  instrumental for state-feedback attitude control designs based on the energy-shaping method.  For more elaborated (for instance,  adaptive control and output (attitude) feedback)  controller designs, the following additional properties will be needed. 

\begin{lem}{\emph{(\textbf{Additional properties of the Lagrangian dynamics \eqref{eq:EL-model}):}}} \label{lem3}
The Lagrangian dynamics \eqref{eq:EL-model} with $D(q)$   in \eqref{eq:Dq2} and $C(q,\dot{q})$  in \eqref{eq:Cqqp} has in addition the following properties:  $\forall x,y,z,v\in\mathbb{R}^{4}$ 
\begin{enumerate}
\item The Lagrangian dynamics \eqref{eq:EL-model} allows a linear parametrization: 
\begin{equation} \label{eq:LP}
D(q)\ddot{q} + C(q,\dot{q})\dot{q} = Y_0(q,\dot{q},\ddot{q})m_{0} + Y(q,\dot{q},\ddot{q})\theta =\bar{\tau}, 
\end{equation}
$\forall q\in\mathcal{S}^{3}$, where 
\begin{align}
    Y_0( q,\dot{q},\ddot{q}) &= \left( q^{T}\ddot{q} + \dot{q}^{T}\dot{q} \right) q \in \mathbb R^4,  \label{eq:RegY0}\\
    Y( q,\dot{q},\ddot{q}) &= J(q)\left( F(\dot{w}) + 2S(w)F(w) \right)\in \mathbb R^{4\times 6}, \label{eq:RegY}   
\end{align}
with $w: = J^{T}(q)\dot{q} \in \mathbb{R}^{3}$, $F(\cdot)$ defined in Appendix \eqref{eq:Freg}, and $\theta$ is a vector defined by the entries of the inertial matrix \eqref{eq:InertialM}
\begin{equation}
\theta =  \left[ m_{11} , m_{22}, m_{33}, m_{23}, m_{13}, m_{12}\right]^{T}\in \mathbb R^6. \label{eq:tht}
\end{equation}
\label{pLinParam}

\item $\dot{D}(q) = C(q,\dot{q}) + C(q,\dot{q})^{T}$ for all $q\in\mathcal{S}^{3}$ that verifies $q^{T}\dot{q} = 0$.
\label{pDC1}
%\item $\|D(x)\|\leq k'_{M}$, $k'_{M} = \bar{m}\|x\|^{2}$, for $\bar{m} \vcentcolon = \lambda_{\mathrm{max}}(M_{0})$ \label{pDC1}

\item $\|\left( D(x)-D(y) \right)v\|\leq k_{M}\|x-y\|\|v\|$, where 
\begin{equation}\label{eq:km}
 k_{M}\vcentcolon = 16\max_{i,j,k,z_{o}}\left\{ \left| \left.\frac{\partial d_{ij}(z) }{\partial z_{k}}\right|_{z=z_{0}} \right|\right\} ,
\end{equation}
for $i,j,k = 1,2,3,4 $, $z_{0}\in\mathbb{R}^{4}$ and $d_{ij}(z)\in \mathbb{R}$ is the $(i,j)$th entry of the matrix $D(z)$.
 \label{pDC2}
\item $C(q,x)$ $=$ $\left[ C^{T}_{1}(q)x\; C^{T}_{2}(q)x \; C^{T}_{3}(q)x \; C^{T}_{4}(q)x \right]^{T}$, where $C^{T}_{k}x$ are column vectors of $C(q,x)$ and $C_{k}(q)\in \mathbb{R}^{4\times 4}$ are continuous matrices in $q$, whose entries  $C_{k_{ij}}(q)\in\mathcal{C}^{\infty}$ are bounded for all $q\in\mathcal{S}^{3}$, $i,j,k = 1,2,3,4$.  \label{pDC3}

\item $\|C(q,x)y\|\leq k_{c_1}\|x\|\|y\|$, with
\begin{equation}\label{eq:kc1}
 k_{c_1}\vcentcolon = 4^{2}\max_{i,j,k,q} \left|C_{k_{ij}}(q)\right| ,
\end{equation}
for any $q\in\mathcal{S}^{3}$, $i,j,k = 1,2,3,4$.
\label{pDC4}

\item $\|C(x,z)v-C(y,u) v\|\leq k_{c_1}\|u-z\|\|v\| + k_{c_2}\|x-y\|\|z\|\|v\|$, where
\begin{equation}\label{eq:kc2}
 k_{c_2}\vcentcolon = 4^{3}\max_{i,j,k,l,q} \left| \frac{\partial C_{k_{ij}}(q) }{\partial q_{l}} \right| ,
\end{equation}
for any $q\in\mathcal{S}^{3}$, where $q_{l}\in [-1,1]$ is the $l$th element of $q$, and $i,j,k = 1,2,3,4$.
\label{pDC5}
\item For any $q_{d},\dot{q}_{d},\ddot{q}_d \in \mathbb{R}^{4}$ bounded, and some $\bar{p}\in\mathbb{R}^{4}$ with $\|\bar{p}\|\leq \rho$, the function 
\begin{align}\label{eq:hres}
    \underbar{h}(t,\Tilde{q},\dot{\Tilde{q}}) &= \left( D(q_{d}) - D(q) \right) \ddot{q}_{d} \notag\\
    &\quad + \left( C(q_{d},\dot{q}_{d}) - C(q,\dot{q}) \right) \dot{q}_{d} \notag\\
    &\quad -\frac{1}{2}\left( Q(q_{d}) - Q(q)\right) \bar{p}
\end{align}
for $\Tilde{q} \vcentcolon = q_{d} - q$, is called the residual dynamics and holds 
\begin{equation}\label{eq:hresZero}
    \underbar{h}(t,0_{4\times 1},0_{4\times 1}) = 0_{4\times 1}
\end{equation}
\begin{equation}\label{eq:hresCond}
    \|\underbar{h}(t,\Tilde{q},\dot{\Tilde{q}})\| \leq k_{h_1}\|\dot{\Tilde{q}}\| + k_{h_{2}}\|\mathrm{Tanh}(\Tilde{q})\|
\end{equation}
\begin{align}
    k_{h_1} &\geq k_{c_1}\|\dot{q}_{d}\| \label{eq:kh1} \\
    k_{h_2} &\geq \frac{s_{2}}{\tanh{(s_{2}/s_{1})}} \label{eq:kh2} \\
    s_{1} &\vcentcolon = 8\rho + k_{M}\|\ddot{q}_{d}\| + k_{c_2}\|\dot{q}_{d}\|^{2} \label{eq:s1} \\
    s_{2} &\vcentcolon = 2 \left( \frac{1}{2}\rho + \bar{m}\|\ddot{q}_{d}\| + k_{c_2}\|\dot{q}_{d}\|^{2}\right) \label{eq:s2}
\end{align}
where $\mathrm{Tanh}(x) \vcentcolon = \left[ \tanh{(x_1)} , \ldots , \tanh{(x_4)} \right]^{T}\in\mathbb R^4$.
\label{pDC6}
\end{enumerate}
\end{lem}
\begin{pf}
See Appendix \ref{AppDCp}.
\begin{flushright}
$\square$
\end{flushright}
\end{pf}

\begin{rem} 
\emph{\textbf{(Derivation of the 4-DOF Lagrangian dynamics):}}
In the  derivation of the 4-DOF Lagrangian dynamics \eqref{eq:EL-model} based on the Euler-Newtonian dynamics  \eqref{eq:Dynamics} 
 the constraint imposed by a unit quaternion $q^{T}q=1$ is incorporated implicitly by restricting the time evolution of $q(t)$ to the kinematic equation \eqref{eq:KinematicsQ}.  The connection between the physically applied torque $\tau$ to the spacecraft and the generalized torque $\bar\tau$ to the quaternion motion  appears apparent through  \eqref{eq:Tau} (cf. \cite{udwadia2010equations} for a detailed discussion).    It is worth mentioning  that in several different but related derivations \cite{sherif2015rotational}, the final description of the Lagrangian dynamics after eliminating the Lagrangian multipliers reaches the same form as \eqref{eq:EL-model}. 
 
% From the controller design point of view, the development of the Lagrangian dynamics presented here is more  straightforward and insightful for exploring the important properties of a Lagrangian system useful for controller design. In fact, the basic energy conservation property of the Lagrangian dynamics \eqref{eq:EL-model} in Lemma \ref{lem2} and  the additional properties in Lemma \ref{lem3} allow to take full advantage of well-developed controller design methodologies for robot manipulators in problems related with the attitude of a rigid body. These properties,  to the best of the authors' knowledge, are unavailable in the literature for attitude dynamics and are instrumental to the controller design in the sequel.

 From the control design perspective, however, the approach followed in the paper is more  straightforward and insightful  for deriving   control-orientated Lagrangian dynamics compared to the approach of modeling the attitude using a Lagrangian function in conjunction with Lagrange  multipliers \cite{udwadia2010alternative,udwadia2010equations,rodriguez2020lagrange}, since it allows to explore the basic energy-conservation property and the additional properties of the Lagrangian dynamics, enabling to take full advantage of well developed control designs based on energy-shaping  methodologies for a broad class of mechanical systems. To the best of the authors' knowledge, these properties are not revealed before in the literature attitude dynamics.
 
 %An alternative method to obtain the Lagrangian dynamics   \eqref{eq:EL-model} is by means of   modeling the attitude by the Euler-Lagrange method for a constrained mechanical system using Lagrangian multipliers \cite{udwadia2010alternative,udwadia2010equations}. 
 
Notice that the artificial inertial parameter $m_0$ in the Lagrangian dynamics \eqref{eq:M0}-\eqref{eq:EL-model} is related with the multiplier in the Euler-Lagrange modelling \cite{udwadia2010alternative,udwadia2010equations}. Since the choice of $m_0$ or the Lagrangian multiplier is not unique, the acceleration of the quaternion should not depend on these parameters. The independence of the Lagrangian multiplier in the Lagrangian dynamics  was  shown in \cite{udwadia2010equations}. The Lagrangian dynamics  \eqref{eq:EL-model} is also independent of $m_0$. In fact, by the Lagrangian dynamics \eqref{eq:EL-model}, the fact of  $D^{-1}(q)=Q(q)M_0^{-1}Q^T (q)$ and $q^T\dot q=0$, and some manipulations it follows that
\begin{align}
   \ddot q &=D^{-1}(q)\Big( -C(q,\dot q)\dot q+\bar \tau\Big) \label{eq:acc} \\  
     &=J(q)M^{-1}S(M\omega)J^{T}(q)\dot q +J(\dot q)J^T(q)\dot q +\frac{1}{2}J(q)M^{-1}\tau \notag
\end{align} 
which results in the same as the acceleration in \eqref{eq:qdd} and is indeed independent of $m_0$. 
\end{rem}

\section{Attitude Tracking Controller Design}\label{Sec:SF}
In this section, four attitude tracking controllers will be designed. Firstly, under ideal conditions, i.e., in the absence  of  measurement noise, torque disturbances, and uncertainty in the inertial matrix, and assuming  both  attitude and angular-velocity measurements are available for feedback, a continuous controller is designed  with global exponential stability. To deal with the ambiguity of two quaternions corresponding to the same attitude, the desired trajectory is "initialized" according to the initial attitude measured by a quaternion by switching it to the same hemisphere  on $\mathcal S^3$   of the initial attitude. To address the issues of attitude  measurement noise, on-orbit torque disturbances and unknown inertial matrix, and lack of angular-velocity measurement for feedback, next a hybrid state-feedback controller, an adaptive hybrid  state-feedback controller, and an adaptive hybrid attitude-feedback  controller are designed.  

\subsection{Control Objectives}
\begin{figure}[tbh!]
	\begin{center}
		\includegraphics[trim = 0mm 0mm 0mm 0mm,scale=0.5]{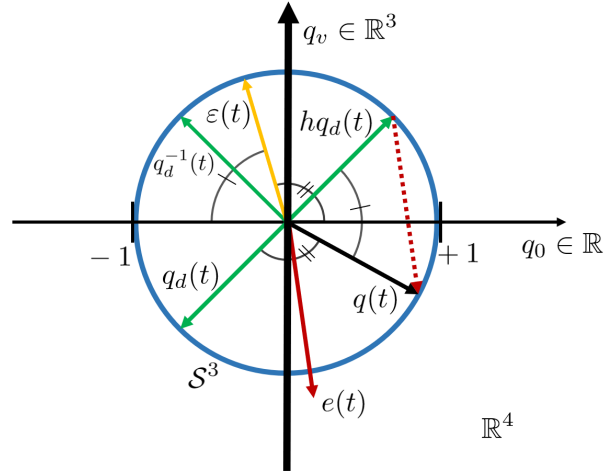}
		\caption{Illustration of the attitude tracking errors: The controller renders the tracking error  $e(t)=q(t)-hq_d(t)\to 0_{4\times 1}$ in the space $\mathbb R^4$, which implies the quaternion error $\varepsilon (t) \to [h \ 0\ 0\ 0]^T$ in the unit sphere $\mathcal S^3$. Given the the initial conditions of $q$ and $q_d$ indicated in the figure, the shortest  path, indicated by the single strip angles, to reach the same physical rotation is between $q$ and the reassigned desired trajectory with the initial condition $h q_d$ with $h=-1$. This implies the quaternion error $\varepsilon$ is driven  to $[-1 \ 0\ 0\ 0]^T$ in the closed loop.}
		\label{fig:TckErrors}
	\end{center}
\end{figure}

Let  $q_d=[ q_{d_0}, q^{T}_{d_v}]^{T} \in\mathcal{S}^{3}$ be a twice differentiable  desired attitude and $\omega_{d}\in\mathbb{R}^{3}$ be the desired angular velocity expressed in the desired reference frame, satisfying 
\begin{equation}
    \dot{q}_{d} = \frac{1}{2}J(q_{d})\omega_{d}, \ \ q_d(0)\in \mathcal{S}^3.  \label{eq:qd}
\end{equation} Define the (Euclidean) attitude tracking error 
\begin{equation}\label{eq:qTilD}
e(t)= q(t) - hq_{d}(t) \in\mathbb R^4
\end{equation}
and the quaternion error 
\begin{equation}\label{eq:epsError}
\varepsilon (t)= Q^T \left(q_{d}\right)q = [\varepsilon_{0} (t), \varepsilon_{v} (t)]^T \in \mathcal{S}^{3},
\end{equation}
where $h$ is a constant for a given initial condition of $\varepsilon_{0} (0)$ defined as
\begin{equation}\label{eq:h}
h\vcentcolon = \widehat{\mathrm{sgn}}(\varepsilon_{0} (0))=
\left\lbrace
\begin{array}{cc}
 1, & \mathrm{if} \  \varepsilon_{0} (0)\geq 0 \\
-1, & \mathrm{if} \  \varepsilon_{0} (0)<0.
\end{array}
\right.
\end{equation}
This parameter is used to reassign the desired trajectory $q_d$ according to the initial condition $\varepsilon_{0} (0)$ to the same hemisphere as the spacecraft quaternion   so that the spacecraft attitude tracks the desired attitude by following the shortest path. Similar reassignment was used in  \cite{li2010global,liu2017robust}.  Figure \ref{fig:TckErrors} illustrate these  attitude errors in the Euclidean space $\mathbb{R}^{4}$. \footnote{The illustration of the unit sphere $\mathcal S^3$ embedded in the space $\mathbb R^4$ is inspired by Fig. 3.10 in \cite{junkins2009analytical}.}

The control objective is then to design a control law for $\bar{\tau}$ to render the attitude tracking error $e \to 0_{4\times 1}$ and $\dot{e}\to 0_{4\times 1}$ globally exponentially stable. This in turns implies that $q\to hq_{d}$ and $\dot{q}\to h\dot{q}_{d}$ globally exponentially. By $h\dot{q}_{d} = \frac{1}{2}J(hq_{d})\omega_{d}$ and Property \ref{pA1} of the matrix $J(\cdot)$, it has $h^{2}\omega_{d} = 2J^{T}(hq_{d})h\dot{q}_{d}$. In consequence,  $q\to hq_{d}$ and $\dot{q}\to h\dot{q}_{d}$ implies $\omega \to h^{2} \omega_{d} = \omega_{d}$. Thus, with a little notation abuse, the desired trajectory  $q_{d}$, $\dot{q}_{d}$ and $\ddot{q}_{d}$ will be used in the ideal state-feedback  control design instead of the reassigned ones $hq_{d}$, $h\dot{q}_{d}$ and $h\ddot{q}_{d}$.

\subsection{State-feedback Controller Design}
Given the desired trajectory $q_d, \dot q_d$, define the reference velocity as 
\begin{equation}\label{eq:qrP}
\dot{q}_{r} = \dot{q}_{d} - \Lambda \left( q - q_{d} \right) 
\end{equation} 
where $0<\Lambda =\Lambda^T \in \mathbb{R}^{4 \times 4}$ is a gain matrix. Define the combined tracking error as 
\begin{equation}\label{eq:s}
s = \dot{q} - \dot{q}_{r} = \dot{q} - \dot{q}_{d} + \Lambda \left( q - q_{d} \right).
\end{equation}
The state-feedback tracking controller is readily  proposed as
\begin{equation}\label{eq:Ctrl}
\bar{\tau} = D(q)\ddot{q}_{r} + C(q,\dot{q})\dot{q}_{r} -K_{s} s,
\end{equation}
where $0<K_{s}=K_s^T \in \mathbb{R}^{4\times 4}$ together with  matrix $\Lambda$ provides the PD-control action through  $K_{s}s = K_{s}\dot{e} + K_{p} e$, with $K_p:=K_{s}\Lambda$.  The actual control torque applied to the spacecraft is calculated by \eqref{eq:Tau}.

\begin{thm}\label{thm1}
\emph{\textbf{(Ideal State-feedback controller):}}
The control law \eqref{eq:Ctrl} in closed loop with the system \eqref{eq:EL-model} drives $q \to q_{d}$ and $\dot{q} \to \dot{q}_{d}$ exponentially from any initial conditions $q(0)\in\mathcal S^3$  and $\dot q (0)\in \mathbb{R}^{4}$.
\end{thm}
\begin{pf}
Taking the time derivative of \eqref{eq:s}, premultiplying this by the matrix $D(q)$ in \eqref{eq:Dq2} and by the Lagrangian dynamics \eqref{eq:EL-model}, gives 
\begin{eqnarray*}
D(q)\dot{s} &=& D(q)(\ddot{q} - \ddot{q}_{r}) \nonumber\\
&=& -C(q,\dot{q})\dot{q} + \bar{\tau} - D(q)\ddot{q}_{r}.
\end{eqnarray*}
This error dynamics in closed loop with the controller  \eqref{eq:Ctrl} results in
\begin{eqnarray}\label{eq:sP}
D(q)\dot{s} &=& -C(q,\dot{q})\dot{q} +C(q,\dot{q})\dot{q}_{r} -K_{s} s,\nonumber\\
&=& -C(q,\dot{q})s - K_{s}s .
\end{eqnarray}

On the other hand,  by \eqref{eq:s} the attitude error dynamics \eqref{eq:qTilD}   is rewritten as
\begin{equation}\label{eq:qTilDp2}
\dot{e} = \dot{q} - \dot{q}_d = s - \Lambda e.
\end{equation}

Notice that the error dynamics \eqref{eq:sP}-\eqref{eq:qTilDp2} has the origin $s=0_{4\times 1}$, $e=0_{4\times 1}$ as the unique equilibrium. Consider the Lyapunov function candidate
\begin{equation}\label{eq:V1}
    V_{1} = \frac{1}{2}s^{T}D(q)s + \frac{\alpha}{2}e^{T}e
\end{equation}
for some $\alpha >0$ that verifies  $\alpha I_{4} < 4K_{s}\Lambda = 4K_{p}$. Clearly, this Lyapunov function candidate is radially unbounded. The time derivative of \eqref{eq:V1} is 
\begin{align*}
    \dot{V}_{1} = s^{T}D(q)\dot{s} + \frac{1}{2} s^{T}\dot{D}(q)s + \alpha e^{T}\dot{e}.
\end{align*}
Its time evolution along the error dynamics \eqref{eq:sP}-\eqref{eq:qTilDp2} is 
\begin{align}\label{eq:V1p}
    \dot{V}_{1} &= s^{T}\left(-C(q,\dot{q})s - K_{s}s \right) + \frac{1}{2} s^{T}\dot{D}(q)s + \alpha e^{T}\left( s - \Lambda e \right) \notag\\
    &= -s^{T}K_{s}s + \alpha e^{T}s - \alpha e^{T}\Lambda e \notag \\
    &= -\left[ 
\begin{array}{cc}
 s^{T} &
 e^{T}
\end{array}
\right]\left[ 
\begin{array}{cc}
K_{s} & -\frac{\alpha}{2} I_{4}  \\
-\frac{\alpha}{2} I_{4} & \alpha\Lambda
\end{array}
\right]\left[ 
\begin{array}{c}
 s \\
 e
\end{array}
\right]
\end{align}
where the skew symmetry  \eqref{pLg2} of the matrix $\dot{D}(q)-2C(q,\dot{q})$ in Lemma \ref{lem2} is used. Thus, \eqref{eq:V1p} is negative definite, concluding that the equilibrium  $s=0_{4\times 1}$, $e=0_{4\times 1}$ is uniformly globally exponentially stable. This implies that $q(t)\to q_d(t)$ and $\dot{q}(t) \to \dot q_d(t)$ exponentially from any initial conditions $q (0)\in \mathcal S^3$, $\dot q(0)\in \mathbb{R}^{4}$. 
\begin{flushright}
$\square$
\end{flushright}
\end{pf}

\begin{rem} \label{rm:rm6}
\emph{\textbf{(State-feedback controller):}} 
The control law \eqref{eq:Ctrl} has been widely studied for tracking control designs for mechanical systems in Euclidean space, particularly for robot manipulators. The proposed 4-DOF Lagrangian approach allows to bridge up  the gap between control designs  for mechanical  systems in Euclidean space and control designs for the attitude of a rigid body in the quaternion group, and to enable using the whole quaternion in attitude control designs, in contrast to control designs  that use only the vector part of  the quaternion  \cite{caccavale1999output,wong2001adaptive},  avoiding the topological obstruction discussed in \cite{bhat2000topological,mayhew2011quaternion} and singularities in controller development using any 3-DOF Lagrangian dynamics to describe the attitude of a rigid body \cite{slotine1990hamiltonian,fossen1991adaptive,tomei1992nonlinear,caccavale1999output,chung2009application}.  The global exponential convergence of the tracking error by the continuous controller \eqref{eq:Ctrl} is achieved on the Euclidean space  $\mathbb R^4$ by designing the control law that renders the origin $(e,\ \dot e)=0$ globally exponentially stable. Notice that $q(t), \ q_{d}(t)\in\mathcal{S}^{3}$ for all $t\geq 0$, so does the quaternion  error $\varepsilon (t)$ defined in \eqref{eq:epsError}.  The convergence to the origin of $e(t)$ implies the convergence of the quaternion error $\varepsilon (t)$ to the quaternion identity in the quaternion group  $\mathcal S^{3}$. 

Compared to the results based on Euler-Newtonian dynamics \cite{egeland1994passivity,lizarralde1996attitude,costic2001quaternion,tayebi2008unit,lee2012exponential,lee2015global,arjun2020uniform}, the proposed controller achieves the global exponential  stability  without the need of addressing the  singularity issues.
\end{rem}

%\section{Discussions and Extensions}  \ref{Sec:ext}

\subsection{Hybrid State-feedback Controller} %State-feedback Hybrid Controller
When the quaternion measurements are clean, the initial assignment of the desired quaternion in \eqref{eq:h} ensures the tracking error $e(t)$ to converge to the origin under the control law \eqref{eq:Ctrl} and \eqref{eq:Tau}. The convergence to the origin of the tracking error $e(t)$ implies convergence of the quaternion error $\epsilon(t)$ in (31) to either of the two disconnected antipodal points on $S^3$. Without a careful treatment of the ambiguity, however, it may become sensitive to small quaternion measurement noise \cite{mayhew2011quaternion}. To deal with this issue, motivated by the recent works \cite{berkane2017hybrid,casau2019robust} this subsection presents a hybrid state-feedback controller.  The main idea is to make the switching as in \eqref{eq:qTilD} each time when a switch condition is satisfied, depending on a potential function $\mathcal{U}(e)$ measuring the closeness to one of these two antipodal points. 

Let $\mathcal{H}=\{-1,1\}\subset \mathbb{N}$ be an index set, and  $\mathcal{U}\vcentcolon \mathbb{R}^{4}\times \mathcal{H} \to \mathbb{R}_{+} $ be a potential function with respect to the set $\mathcal{A}_{0}= \{0_{4\times 1}\} \times \mathcal{H}$. That is, for each $h\in\mathcal{H}$, the map $e\mapsto \mathcal{U}(e,h)$ is continuously differentiable in $\mathcal{D}\subseteq \mathbb{R}^{4}\times \mathcal{H}$.
%, $\forall (e,h)\in \mathcal{D}$. 
In addition, $\mathcal{U}(e,h)$ verifies $\mathcal{U}(e,h)>0$ for all $(e,h)\notin \mathcal{A}_{0} $, $\mathcal{U}(e,h)=0$ for all $(e,h)\in \mathcal{A}_{0} $, and
\begin{equation}\label{eq:gap}
   \mathcal{G}(e)\vcentcolon = \mathcal{U}(e,h) - \underset{m\in\mathcal{H}}{\min} \;\mathcal{U}(e,m) >\delta,
\end{equation}
for some $\delta > 0$. Then, inspired by the switching mechanism of \cite{lee2015global} the discrete state $h$, which dictates the current mode of the hybrid control system, is defined as

\begin{equation}\label{eq:hfcn}
    \left\{ \begin{array}{cc}
         \dot{h} = 0,\quad\quad\quad\quad\quad\quad\quad & \;\;\mathrm{if} \ \mathcal{G}(e) \leq \delta ,  \\
         h^{+}\in\arg\underset{m\in\mathcal{H}}{\min} \;\mathcal{U}(e,m), & \;\; \mathrm{if} \ \mathcal{G}(e) \geq \delta .
    \end{array}\right.
\end{equation}

For the design of the hybrid controller, the potential function $\mathcal{U}(e,h) \vcentcolon = \|e\|^{2}$ is considered. Observe that its gradient with respect to $e$ is  $\nabla \mathcal{U} = 2e$, which vanishes only at $\mathcal{A}_{0}$.
 
Let the set of critical points for $\mathcal{U}(e,h)$ be defined as $\Omega_{0} = \{ (e,h)\;|\; e = 0_{4\times 1} \}$ for $(e,h)\in\mathcal{D}$. Note that all potential functions $e\mapsto \mathcal{U}(e,h)$, $\forall h\in \mathcal H$, share the same critical point $e= 0_{4\times 1}$, i.e., the potential function $\mathcal{U}$ is centrally synergistic with a gap exceeding $\delta > 0$ \cite{berkane2017hybrid,mayhew2013synergistic}. 

The proposed hybrid controller consists of \eqref{eq:Tau},
\begin{align}
    \bar{\tau} &= D(q)\ddot{q}_{r} + C(q,\dot{q})\dot{q}_{r} - K_{s}s(\dot{e},e,h), \label{eq:CtrlHyb} \\
    \dot{q}_{r} &= \dot{q}_{d} - \Lambda e, \label{eq:hybqr}\\
    s &= \dot{e} + \Lambda e, \label{eq:shyb} \\
    e &= q-hq_{d}, \label{eq:ehyb} \\
    \dot{e} &= \dot{q}-h\dot{q}_{d}, \label{eq:ephyb}
\end{align}
and the switching mechanism \eqref{eq:hfcn}.

\begin{thm}\label{clr1}
\emph{\textbf{(Hybrid state-feedback controller):}}
The hybrid controller \eqref{eq:Tau}, \eqref{eq:CtrlHyb}-\eqref{eq:ephyb},  and \eqref{eq:hfcn} in closed loop with the system \eqref{eq:EL-model} stabilizes the set $\mathcal{A}_{1}:= \{ (e,s,h)\in\mathcal{X}_{1} |  e=0_{4\times 1}, s = 0_{4\times 1} \}$ globally exponentially, where $\mathcal{X}_{1}\vcentcolon= \mathbb{R}^{4}\times  \mathbb{R}^{4}\times\mathcal{H}$. 
\end{thm} %\end{cor}

\begin{pf}
Define $\zeta_{1} \vcentcolon = \left( e,s, h \right) \in\mathcal{X}_{1}$. Then, the closed-loop system is described by
\begin{equation} \label{eq:ClsdHyb}
    \left. \begin{array}{rl}
        \dot{e} =& s - \Lambda  e \\
        D(q)\dot{s} =& -C(q,\dot{q})s - K_{s}s\\
        \dot{h} =& 0
    \end{array}
    \right\} \quad\forall\zeta_{1}\in\bar{C}_{1} , 
\end{equation}
\begin{equation} \label{eq:ClsdHyb1}
        \zeta_{1}^{+} = \left[ \begin{array}{c}
             e \\
             s \\
             h^+
        \end{array} \right] ,  \quad\forall\zeta\in\bar{D}_{1} ,
\end{equation}
where $\zeta_{1}^+$ is the state value right after the jump, $h^+$ is given in \eqref{eq:hfcn}, and $\bar{C}_{1}$ and $\bar{D}_{1}$ represent the flow set and the jump set, respectively, given by
\begin{align}
 \bar{C}_{1} &= \{ \zeta_{1}\in\mathcal{X}_{1}   \ | \ \mathcal{G}(e) \leq \delta \} , \label{eq:SetC} \\
  \bar{D}_{1} &= \{ \zeta_{1}\in\mathcal{X}_{1}   \; | \;  \mathcal{G}(e) \geq \delta \} . \label{eq:SetD}
\end{align}

Consider the following Lyapunov function candidate
\begin{equation}\label{eq:V2}
    V_{2}(\zeta_{1}) = \frac{1}{2}s^{T}D(q)s + \frac{\alpha}{2}\mathcal{U}(e,h),
\end{equation}
for some $0 < \alpha < 4\lambda_{\min}(K_{s})\lambda_{\min}(\Lambda)$. Note that $V_{2}$ is bounded by $ \frac{1}{2}z^{T}M_{1}z \leq V_{2} \leq \frac{1}{2}z^{T}M_{2}z$, with $z=\left[ \|s\| , \|e\| \right]^{T}$, and
\begin{equation*}
  M_{1} = \left[ \begin{array}{cc}
         \underbar{m} & 0  \\
         0 & \frac{\alpha}{2} 
    \end{array}\right] , \quad     M_{2} = \left[ \begin{array}{cc}
         \bar{m} & 0  \\
         0 & 2\alpha 
    \end{array}\right] .
\end{equation*}

Then, in view of \eqref{eq:V1p} and \eqref{eq:hfcn} the time derivative of $V_{2}$ along \eqref{eq:ClsdHyb} for all $\zeta_{1} \in \bar{C}_{1}$ is given by
\begin{align}\label{eq:V2p}
    \dot{V}_{2} &= -s^{T}K_{s}s +\alpha e^{T}s -\alpha e^{T}\Lambda e, \notag\\
    &\leq -z^{T}\left[ \begin{array}{cc}
         \lambda_{\min}(K_{s}) & -\frac{\alpha}{2}  \\
         -\frac{\alpha}{2}& \alpha\lambda_{\min}(\Lambda) 
    \end{array}\right] z \notag \\
    &=  -z^{T}M_{3}z \leq -\lambda_{c}V_{2},
\end{align}
where $\lambda_{c} \vcentcolon = \frac{\lambda_{\min}(M_{3})}{\lambda_{\max}(M_{2})} >0 $. Therefore, the states $s,e\in\mathbb{R}^{4}$ are bounded for any $h\in\mathcal{H}$, and $\zeta_{1}\in \bar{C}_{1}$.

In the jump set  $\zeta_{1} \in \bar{D}_{1}$, it yields  
\begin{align}\label{eq:JmpSetDiff}
    V_{2}(\zeta_{1}^{+}) - V_{2}(\zeta_{1}) &= \frac{1}{2}s^{T}D(q)s + \frac{\alpha}{2}\mathcal{U}(e,h^+) \notag \\
    &\quad - \frac{1}{2}s^{T}D(q)s -  \frac{\alpha}{2}\mathcal{U}(e,h) \notag\\
    & \leq -\frac{\alpha}{2}\delta , 
\end{align}
therefore, for $\alpha>1$, $V_{2}$ is strictly decreasing over the jump set $\bar{D}_{1}$. Thus, $\zeta_{1}(t)$ is bounded for all $t\geq 0$. 
Similar to the proof of Theorem 2 in \cite{berkane2017hybrid},  it  follows that $V_{2}(\zeta_{1}(t,j)) \leq \mathrm{e}^{-\lambda_{2}(t+j)}V_{2}(\zeta_{1}(0,0))$ for all $(t,j)\in \mathrm{dom}\zeta_{1}$, where $\lambda_{2} = \min \left\{ \lambda_{c},  -\ln{\left( 1- \frac{\alpha \delta}{4\bar{v}}\right)} \right\}$, given $V_{2}\leq \bar{v}$. Therefore, the equilibrium point $(s,e) = (0_{4\times 1}, 0_{4\times 1} )$ is globally exponentially stable.
\begin{flushright}
$\square$
\end{flushright}
\end{pf}

\begin{figure}[tbh!]
	\begin{center}
		\includegraphics[trim = 0mm 0mm 0mm 0mm,scale=0.85]{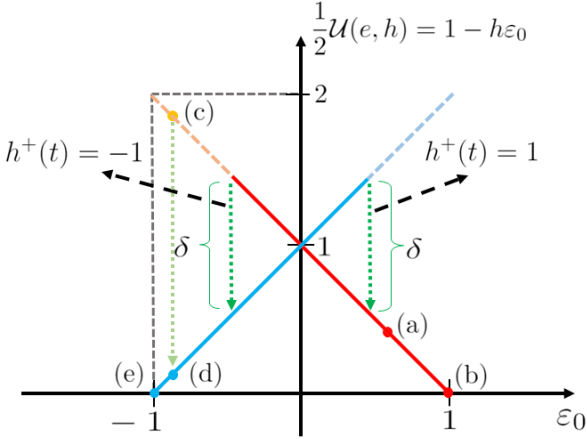}
		\caption{Illustration of the potential function. The potential function $\mathcal{U}(e,h)$ is represented by the red and blue solid lines corresponding to  $\mathcal{U}(e,1)$ and $\mathcal{U}(e,-1)$, respectively. They are extended by the dashed lines corresponding to the switching zone, i.e., where $\mathcal{U}(e,h) - \underset{m\in\mathcal{H}}{\min} \;\mathcal{U}(e,m) \geq \delta$. Any trajectory starting in the red solid line under the continuous controller of Theorem \ref{thm1}, for example (a), is led to $1$ (b). On the other hand, any point in the blue solid line (d) is led to $-1$ (e). Notice that under a noisy condition a trajectory starting at (a), which would converge to (b) in the noise-free condition, can be led to (c) under a "larger"     noise level, where $\mathcal{U}(e,1) - \mathcal{U}(e,-1) \geq \delta$. Given that $\underset{m\in\mathcal{H}}{\min} \;\mathcal{U}(e,m) = \mathcal{U}(e,-1)$ in (c), the hybrid controller in Corollary \ref{clr1} performs a jump $h^{+}=1$, to (d). Consequently, this trajectory eventually converges to (e). %Therefore, this potential function allows to stabilize  $\varepsilon_{0} = \pm 1$ by the shortest path.
		}
		\label{fig:PFcn}
	\end{center}
\end{figure}

\begin{rem} \label{rm:rm8}
\emph{\textbf{(The centrally synergistic potential function $\mathcal{U}(e,h)$):}}
The potential function $\mathcal{U}(e,h) = \|e\|^{2}$, for $e=q-hq_{d}$ can be expressed in terms of the scalar part $\varepsilon_0$ of the quaternion error $\varepsilon$ \eqref{eq:epsError}  as $\mathcal{U}(e(\varepsilon_0),h) = 2(1-h\varepsilon_{0}) :=\mathcal U_\varepsilon (\varepsilon_0,h)$ (shown in Fig.  \ref{fig:PFcn}). This potential function has been widely used to globally stabilize the set $\mathcal A = \{ \pm \hat{1} \}\times \mathcal{H}$, where $\hat{1} = [1,0,0,0]^{T}\in\mathcal{S}^{3}$, see for instance \cite{mayhew2011synergistic,mayhew2011quaternion,wisniewski2003rotational}. However, the potential function $\mathcal{U}_\varepsilon(\varepsilon_0,h)$ relative to $\mathcal A$ is not central \footnote{The reader is referred  to \cite{mayhew2011synergistic,berkane2017hybrid} for more details.}, i.e., the potential functions $\varepsilon\mapsto \mathcal U_\varepsilon (\varepsilon_0,h)$, for $h\in\mathcal H$, have different critical points $\pm\hat{1}$. On the other hand, the potential function $\mathcal{U}(e,h) = \|e\|^{2}$ relative to $\mathcal{A}_{0}$ is central, for having the same critical point $e=0_{4\times 1}$ for $h\in \mathcal H$, which guarantees that each mode of the hybrid closed-loop system is almost globally exponentially stable \cite{berkane2017hybrid}.  

Notice that unlike the potential function defined in a manifold, e.g., $SO(3)$ or $\mathcal S^3$ \cite{berkane2017hybrid,casau2019robust}, the potential function $\mathcal U_\varepsilon(\varepsilon_0,h)$ has a unique feature of containing no undesired critical points since both $\varepsilon_0=\pm 1$ achieve the attitude tracking $R=R_d$. On the other hand, since the controller design is carried out on $\mathbb R^4$, the potential function $\mathcal U(e,h)$ has only one critical point, which corresponds to the desired equilibrium. 
\end{rem}

\subsection{Adaptive Hybrid State-feedback Controller}
This subsection addresses the issue of parameter uncertainty of the inertia matrix $\theta\in\mathbb{R}^{6}$ in \eqref{eq:tht} and on-orbit disturbance torque, assumed to be an unknown constant vector. This assumption is made for simplicity and it is a good model for on-orbit torque disturbances as justified in \cite{lee2015global}. Furthermore, the adaptive control law given in this subsection can track slowly time-varying torque disturbances. 

The Lagrangian system \eqref{eq:EL-model} in the presence of disturbances $p\in\mathbb{R}^3$ can be expressed as
\begin{equation}\label{eq:EL-modelD}
    D(q)\ddot{q} + C(q,\dot{q})\dot{q} = \bar{\tau} + d,
    %= \frac{1}{2}J(q)(\tau + p),
\end{equation}
where $d\in\mathbb{R}^{4}$ is   % the generalized torque disturbance
\begin{equation}\label{eq:dDist}
    d = \frac{1}{2}J(q)p = \frac{1}{2}Q(q)\bar{p},
\end{equation}
with $\bar{p}\vcentcolon = \left[0, \ p^{T}\right]^{T}\in\mathbb{R}^{4}$ representing the unknown constant disturbance. By the linear parameterization \eqref{eq:LP} in Lemma \ref{lem3}, the dynamics \eqref{eq:EL-modelD} can be rewritten as
\begin{align}\label{eq:LparamEL}
    Y_{0}(q,\dot{q},\ddot{q})m_{0} + \bar{Y}(q,\dot{q},\ddot{q})\Theta &= \bar{\tau},
\end{align}
where $\Theta \vcentcolon = \left[ \theta^{T}, \  p^{T} \right] ^{T}\in\mathbb{R}^{9}$ is the augmented vector of unknown parameters. The regressor $\bar{Y}(q,\dot{q},\ddot{q})\in\mathbb{R}^{4\times 9}$ is defined as 
\begin{equation}\label{eq:Ybar}
\bar{Y}(q,\dot{q},\ddot{q}) = \left[ Y(q,\dot{q},\ddot{q}) \ -\frac{1}{2}J(q) \right].
\end{equation}

Define the parameter error
\begin{equation}\label{eq:ThtTilD}
    \tilde{\Theta} = \hat{\Theta} - \Theta,
\end{equation}
with  $\hat{\Theta}\in\mathbb{R}^{9}$ the parameter estimate, and an auxiliary variable $\eta_{1}\in\mathbb{R}^{4}$
\begin{equation}\label{eq:eta1}
    \eta_{1} = \dot{e} + e,
\end{equation}
with  $\dot{e},e\in\mathbb{R}^{4}$ given by \eqref{eq:ehyb} and \eqref{eq:ephyb}, respectively.  
 
Let the state be $\zeta_{2}= \left( e,\eta_{1},\tilde{\Theta},h \right) \in \mathcal{X}_{2}\vcentcolon = \mathbb{R}^{4}\times  \mathbb{R}^{4}\times\mathbb{R}^{9}\times\mathcal{H}$, where the discrete state $h\in\mathcal{H}$ defined in \eqref{eq:hfcn}. Then, the flow set and the jump set are given by
\begin{align}
 \bar{C}_{2} &= \{ \zeta_{2}\in\mathcal{X}_{2}   \ | \ \mathcal{G}(e) \leq \delta \}, \label{eq:SetbarC2} \\
  \bar{D}_{2} &= \{ \zeta_{2}\in\mathcal{X}_{2}   \; | \;  \mathcal{G}(e) \geq \delta \} . \label{eq:SetbarD2}
\end{align}

The following adaptive state-feedback controller is proposed
\begin{equation}\label{eq:AdpCtrl}
    \bar{\tau} = Y_{d_0}m_{0} + \bar{Y}_{d}\hat{\Theta} - K_{d}\eta_{1} - k_{p}e,
\end{equation}
where $0< K_{d}^{T}=K_{d}\in\mathbb{R}^{4\times 4}$ and, $k_{p}>0$ are design parameters. 
The feedforward compensation is given by $Y_{d_0}m_{0} + \bar{Y}_{d}\hat{\Theta}=D(q_{d})\ddot{q}_{d} + C(q_{d},\dot{q}_{d})\dot{q}_{d}$, with $Y_{d_{0}}\vcentcolon = Y_{0}(q_{d},\dot{q}_{d},\ddot{q}_{d})$ and  $\bar{Y}_{d} \vcentcolon = \bar{Y}(q_{d},\dot{q}_{d},\ddot{q}_{d})$ given in  \eqref{eq:RegY0} and \eqref{eq:Ybar}, respectively.

The parameter estimate $\hat{\Theta}$ is updated according to the following adaptation law
\begin{align}
    \dot{\hat{\Theta}} &= - \gamma_{1} Y^{T}_{f}\left( \hat{\tau}_{f}  - \tau_{f}\right) - \gamma_{2}\bar{Y}^{T}_{d}\eta_{1} ,  \label{eq:ThtEstp} \\
    \hat{\tau}_{f} &= Y_{f}\hat{\Theta},     \label{eq:ThtEstp1}
\end{align}   
with $\gamma_{1}>0$, $\gamma_{2}>0$ design parameters. Additionally,  matrix $Y_{f}(q,\dot{q},q_{f})\in\mathbb{R}^{4\times 9}$, and  vectors $\tau_{f}\in\mathbb{R}^{4}$, 
$q_{f}\in\mathbb{R}^{4}$ are obtained by a first-order linear filter, aimed at avoiding  $\ddot q$ to appear in the regressors, as follows
\begin{align}
    Y_{f}(q,\dot{q},q_{f}) &= \left[ \lambda_{f}J(q)F(w)-X_{f} \;\; -\frac{1}{2}J(q_{f}) \right] ,\label{eq:Yf} \\
    \dot{X}_{f} &= \lambda_{f}\left( X - X_{f}\right) , \quad X_{f}(0) = X(0), \label{eq:Xp} \\
    X &\vcentcolon = \lambda_{f}J(q)F(w) - 2J(q)S(w)F(w) \notag\\
    &\quad +J(\dot{q})F(w) ,\label{eq:Xmat} \\
    \dot{\tau}_{f} &= \lambda_{f} \left( \bar{\tau} - \tau_{f}\right), \quad \tau_{f}(0) = \bar{\tau}(0), \label{eq:taufp}\\
    \dot{q}_{f} &= \lambda_{f} \left( q - q_{f}\right), \quad q_{f}(0) = q(0), \label{eq:qfp}
\end{align}
with $\lambda_{f}>0$ the filter gain. Map $F(\cdot)$ and  vector $w\vcentcolon = J^{T}(q)\dot{q}\in\mathbb{R}^{3}$ are given in Lemma \ref{lem3}.

\begin{thm}\label{thm2}
\emph{\textbf{(Adaptive hybrid state-feedback controller):}}
Choose the design parameters $k_{p}>0$, $\gamma_{1}>0$, $\gamma_{2}>0$, $\lambda_{f}>0$, $m_{0}>0$, and matrix $K_{d} = K^{T}_{d} >0$ such that
\begin{align}
    \lambda_{\min}\left( K_{d}\right) &> \frac{(\alpha_{1}+\alpha_{2})^{2}}{4k_{p}} + \alpha_{1}, \label{eq:AdpCtrlCond1}
\end{align}
where
\begin{align}
\alpha_{1} &= k_{h_1} + 2 k_{c_1} + \bar{m}, \label{eq:alph1}\\
\alpha_{2} &= k_{h_{2}} + k_{c_1}\|\dot{q}_{d}\|. \label{eq:alph2}
\end{align}

Then, the hybrid adaptive state-feedback control law \eqref{eq:AdpCtrl} and \eqref{eq:ThtEstp}-\eqref{eq:qfp} in closed loop with the system \eqref{eq:EL-modelD} stabilizes the set $\mathcal{A}_{2} \vcentcolon = \{ (e,\eta_{1} , h)\in\mathbb{R}^{4}\times \mathbb{R}^{4}\times \mathcal{H} |  e= 0_{4\times 1}, \eta_{1}=0_{4\times 1} \}$ globally asymptotically, while  $\tilde{\Theta}$ is maintained  bounded.
\end{thm}

\begin{pf}
Notice that dynamics \eqref{eq:EL-modelD} can be expressed, in view of Property \ref{pDC1} in Lemma \ref{lem3}, as
\begin{equation}\label{eq:EL-modelD2}
    \frac{d}{dt}\left(D(q)\dot{q}\right) - C(q,\dot{q})^{T}\dot{q} - d = \bar{\tau}.
\end{equation}

Therefore, by   \eqref{eq:EL-modelD2} and \eqref{eq:Yf} the filtered torque $\tau_{f}$ obtained from \eqref{eq:taufp} is expressed  as
\begin{equation}\label{eq:tauf}
    \tau_{f} = Y_{f}\Theta.
\end{equation}

Taking the time derivative of the parameter  error \eqref{eq:ThtTilD} and substituting \eqref{eq:ThtEstp} and \eqref{eq:tauf},  yields
\begin{equation}\label{eq:ThtTilDp}
    \dot{\tilde{\Theta}} = -\gamma_{1}Y^{T}_{f}Y_{f}\tilde{\Theta} - \gamma_{2}\bar{Y}^{T}_{d}\eta_{1}.
\end{equation}
The dynamics of $\eta_1$ is obtained by premultiplying the inertia matrix $D(q)$ with the time derivative of \eqref{eq:eta1} as
\begin{align} 
    D(q)\dot{\eta}_{1} &= D(q)\ddot{q} - D(q)\ddot{q}_{d} + D(q)\dot{e}, \notag\\
    &= -C(q,\dot{q})\dot{q} +\bar{\tau} + d - D(q)\ddot{q}_{d} + D(q)\dot{e},  \notag
\end{align}
where \eqref{eq:EL-modelD} is used. Substituting the  controller \eqref{eq:AdpCtrl} into this last equation and after some arrangement yields
\begin{align}\label{eq:eta1p}
    D(q)\dot{\eta}_{1} &= -C(q,\dot{q})\eta_{1} -K_{d}\eta_{1} - k_{p}e + \bar{Y}_{d}\tilde{\Theta} + \chi_{1},
\end{align}
where $\chi_{1} \in\mathbb{R}^{4}$ is defined as
\begin{equation}\label{eq:chi1}
    \chi_{1} = \underbar{h}(t,e,\dot{e}) + C(q,\dot{q})e + D(q)\dot{e},
\end{equation}
with $\underbar{h}(t,e,\dot{e})$ the residual dynamics defined in \eqref{eq:hres}. By Property \ref{pDC6} in Lemma \ref{lem3},  $\chi_{1}$ can be upper bounded by
\begin{equation}\label{eq:UppChi1}
    \|\chi_{1}\|\leq \alpha_{1}\|\eta_{1}\| + (\alpha_{1}+\alpha_{2})\|e\|,
\end{equation}
with $\alpha_{1}$ and $\alpha_{2}$  given in \eqref{eq:alph1}-\eqref{eq:alph2}.

Therefore, the closed-loop system is given by
\begin{equation}\label{eq:ClsLoopSysAdp1C}
  \underbrace{\begin{array}{rl}
        \dot{e} =& -e + \eta_{1} \\
        D(q)\dot{\eta}_{1} =& -C(q,\dot{q})\eta_{1} -K_{d}\eta_{1} - k_{p}e + \bar{Y}_{d}\tilde{\Theta} + \chi_{1}\\
        \dot{\tilde{\Theta}} =& -\gamma_{1}Y^{T}_{f}Y_{f}\tilde{\Theta} - \gamma_{2}\bar{Y}^{T}_{d}\eta_{1} \\ 
        \dot{h} =& 0
    \end{array} }_{ \forall \zeta_{2}\in\bar{C}_{2}}
\end{equation}
\begin{equation}\label{eq:ClsLoopSysAdp1J}
    \zeta^{+}_{2} = \left[ \begin{array}{c}
         e \\
         \eta_{1} \\
         \tilde{\Theta} \\
         h^{+}
    \end{array} \right] , \quad \forall \zeta_{2}\in\bar{D}_{2}.
\end{equation}

Let $V_{3}(\zeta_{2})$ be a Lyapunov function candidate defined as
\begin{equation}\label{eq:V3}
    V_{3}(\zeta_{2}) = \frac{1}{2}\eta^{T}_{1}D(q)\eta_{1} +\frac{k_{p}}{2}\mathcal{U}(e,h) + \frac{1}{2\gamma_{2}}\tilde{\Theta}^{T}\tilde{\Theta}.
\end{equation}

Then, $V_{3}$ is bounded by $\frac{1}{2} z^{T}M_{3}z  \leq V_{3} \leq \frac{1}{2} z^{T}M_{4}z$, where $z=\left[ \|\eta_{1}\|,\|e\|,\|\tilde{\Theta}\| \right]$ and 
\begin{equation*}
    M_{3} = \left[ \begin{array}{ccc}
         \underbar{m} & 0 & 0\\
           0 & \frac{k_{p}}{2} & 0 \\
           0 & 0 & \frac{1}{2\gamma_{2}}
    \end{array} \right] , \quad M_{4} = \left[ \begin{array}{ccc}
         \bar{m} & 0 & 0\\
           0 & k_{p} & 0 \\
           0 & 0 & \frac{1}{\gamma_{2}}
    \end{array} \right] .
\end{equation*}

The time derivative of \eqref{eq:V3} along \eqref{eq:ClsLoopSysAdp1C} for all $\zeta_{2}\in\bar{C}_{2}$ is 
\begin{align}\label{eq:V3p}
    \dot{V}_{3} &= \eta^{T}_{1}D(q)\dot{\eta}_{1} + \frac{1}{2}\eta^{T}_{1}\dot{D}(q)\eta_{1}  + k_{p}e^{T}\dot{e} + \frac{1}{\gamma_{2}}\tilde{\Theta}^{T}\dot{\tilde{\Theta}} \notag\\
    &= \eta^{T}_{1} \left( -C(q,\dot{q})\eta_{1} -K_{d}\eta_{1} - k_{p}e + \bar{Y}_{d}\tilde{\Theta} + \chi_{1}\right) \notag \\
    &\quad + \frac{1}{2}\eta^{T}_{1}\dot{D}(q)\eta_{1}  + k_{p}e^{T}\left( -e + \eta_{1} \right)  \notag\\
    &\quad + \frac{1}{\gamma_{2}}\tilde{\Theta}^{T}\left( -\gamma_{1}Y^{T}_{f}Y_{f}\tilde{\Theta}  -\gamma_{2}\bar{Y}^{T}_{d}\eta_{1} \right) \notag\\
    &= -\eta^{T}_{1}K_{d}\eta_{1} - k_{p}e^{T}e + \eta^{T}_{1}\chi_{1} - \frac{\gamma_{1}}{\gamma_{2}}\tilde{\Theta}^{T} Y^{T}_{f}Y_{f} \tilde{\Theta} \notag \\
    &\leq -\lambda_{\min}(K_{d})\|\eta_{1}\|^{2} - k_{p}\|e\|^{2} + \alpha_{1}\|\eta_{1}\|^{2} \notag\\
    &\quad  +(\alpha_{1}+\alpha_{2})\|\eta_{1}\|\|e\| -\frac{\gamma_{1}}{\gamma_{2}}\lambda_{\min}\left(Y^{T}_{f}Y_{f}\right) \|\tilde{\Theta}\|^{2} \notag \\
    &= -z^{T}\Phi_{1} z,
\end{align}
where \eqref{eq:UppChi1} and the skew-symmetric property \eqref{pLg2} of Lemma \ref{lem2} are used. The matrix $\Phi_{1}\in\mathbb{R}^{3\times 3}$ in \eqref{eq:V3p} is given by
\begin{equation}\label{eq:Phi1}
    \Phi_{1} = \left[ \begin{array}{ccc}
         \lambda_{\min}(K_{d}) - \alpha_{1} & -\frac{(\alpha_{1}+\alpha_{2})}{2} & 0\\
         -\frac{(\alpha_{1}+\alpha_{2})}{2}& k_{p}  & 0\\
         0&0 & \frac{\gamma_{1}}{\gamma_{2}}\lambda_{\min}\left(Y^{T}_{f}Y_{f}\right)
    \end{array} \right],
\end{equation}
which is,  under condition \eqref{eq:AdpCtrlCond1}, semi-negative definite. Therefore, state $\eta_{1},e \in\mathbb{R}^{4}$ and $\tilde{\Theta}\in\mathbb{R}^{9}$ remain bounded for all $\zeta_{2}\in\bar{C}_{2}$. On the other hand, noticing that neither $\eta_{2}$ nor $\tilde{\Theta}$ change over the jumps,  it has for $\zeta_{2}\in\bar{D}_{2}$ that
\begin{align}\label{eq:diffJumpSetAdp1}
    V_{3}(\zeta^{+}_{2}) - V_{3}(\zeta_{2}) &= \frac{k_{p}}{2}\mathcal{U}(e,h^{+}) - \frac{k_{p}}{2}\mathcal{U}(e,h) \notag\\
    &\leq -\frac{k_{p}}{2}\delta .
\end{align} 

Thus, $V_{3}(\zeta_{2})$ is  non-increasing  under condition \eqref{eq:AdpCtrlCond1} along trajectories of the closed-loop system  for all $\zeta_{2}\in\bar{C}_{2}$, and  strictly decreasing over the jump set $\bar{D}_{2}$ provided that $k_{p}>0$. Therefore, the set $\mathcal{A}_{2}$ is uniformly globally stable. 

Denote $t_{j}>0$  the time of the $j$th jump for some $j\in\mathbb{N}$. Then, in view of \eqref{eq:diffJumpSetAdp1} and using the same arguments as in the proof of Theorem 3.1 in \cite{gui2016global}, it can be expressed $V_{3}(\zeta^{+}_{2}(t_{j})) - V_{3}(\zeta_{2}(0)) < -j\sigma$, with $\sigma \vcentcolon = \frac{k_{p}}{2}\delta > 0$. Therefore,  for any $V_{3}(\zeta_{2}(t_{j})) >0$ it yields $j < V_{3}(\zeta_{2}(0)) / \sigma$, which implies that the number of jumps is finite. Then, assuming $\zeta_{2}\in\bar{C}_{2}$, and relying on the fact that the states $\eta_{1},e \in\mathbb{R}^{4}$ and $\tilde{\Theta}\in\mathbb{R}^{9}$ are bounded, it is straightforward to verify that $\ddot{V}_{3}$ is bounded.  Thus, by invoking the Barbalat's Lemma, it concludes  that $\dot{V}_{3}\to 0$ as $t\to\infty$. In consequence, the set $\mathcal{A}_{2}$ is uniformly globally asymptotically stable.
\begin{flushright}
$\square$
\end{flushright}
\end{pf}

\begin{rem} \label{rm:newrm}
The adaptation law \eqref{eq:ThtEstp}-\eqref{eq:ThtEstp1} in its analysis form \eqref{eq:ThtTilDp} coincides with the so-called composite adaptation law used for robot control (Eq. (8.125), \cite{slotine1991applied}). It has a smother and faster convergence behavior due to its first-order dynamics  in \eqref{eq:ThtTilDp}. If, in addition, the following persistent excitation
\begin{equation}\label{eq:PEYf}
    \int^{t+T}_{t}Y^{T}_{f}(\uptau)Y_{f}(\uptau) d\uptau \geq \lambda_{z}I_{9}, \quad \forall t\geq 0
\end{equation}
is fulfilled for  the matrix $Y_{f}(t):=Y_f(q(t),\dot q(t),q_f(t)) \in\mathbb{R}^{4\times 9}$ defined in \eqref{eq:Yf} for some $T>0$ and $\lambda_{z}>0$, then  it can be shown using the standard arguments of adaptive control for a linearly parametrized uncertainty  (e.g., \cite{slotine1991applied})  
that the set $\{ \zeta_{2}\in\mathcal{X}_{2} | e= 0_{4\times 1}, \eta_{1}=0_{4\times 1}, \tilde{\Theta} = 0_{9\times 1} \}$  is  globally exponentially stable. 
\end{rem}

\subsection{Adaptive Hybrid Attitude-feedback Controller} \label{Sec:OF-Adp}
In this subsection, the issues of uncertainty in the inertial matrix, constant torque disturbances,  and lack of angular-velocity measurements for feedback  due to, for instance, gyros sensor failures, and unknown inertia matrix are addressed.
The damping term required in the control action is provided by a nonlinear filter, similar to that proposed in \cite{costic2001quaternion}.

Let the  auxiliary variable $\eta_{2}$ be defined as  
\begin{equation}\label{eq:Eta2}
    \eta_{2} = \dot{e} + e + \nu
\end{equation}
where $e$ and $\dot{e}$ are defined in \eqref{eq:ehyb} and \eqref{eq:ephyb}, respectively. 
The required damping is achieved through the nonlinear filter 
\begin{align}
    \dot{e}_{f} &= -\mathrm{Cosh}^{2}(e_{f})\left( K_{f}\nu + k_{v}\eta_{2} - k_{p}e \right),   \label{eq:ef} \\
    \nu &= \mathrm{Tanh}(e_{f}),  \label{eq:nu}
\end{align}
with the initial condition $e_{f}(0)=0_{4\times 1}$, where the hyperbolic functions are entry-wise defined vectors and matrix, respectively
\begin{align*}
\mathrm{Tanh}(e_{f})&:=[\tanh{(e_{f_1})}, ..., \tanh{(e_{f_4})} ]^{T}\in \mathbb R^4\\
\mathrm{Cosh}(e_f)&:= \mathrm{diag}\lbrace \cosh{(e_{f_1})},\ldots ,\cosh{(e_{f_4})} \rbrace \in \mathbb{R}^{4\times 4}, 
\end{align*}
the gain $K_{f}\in\mathbb{R}^{4\times 4}$ is a positive definite symmetric matrix and, $k_{p}>0$ and $k_{v}>0$ are positive constants.

Note that, according to \eqref{eq:nu}, it has  
\begin{equation}\label{eq:nuUp}
    \|\nu\|=\| \mathrm{Tanh}(e_{f})\| \leq 2.
\end{equation}

Implementation of the nonlinear filter \eqref{eq:nu} requires  $\dot e$ involved in the auxiliary variable $\eta_{2}$, which can be eliminated in  a similar way to that of \cite{zhang2000global}:
\begin{align}
    \dot{g} &= -K_{f}\left( g - k_{v}e \right) - k_{v}\left( g + (1-k_{v})e \right) + k_{p}e  \label{eq:pP}\\
    \nu &= g -k_{v}e \label{eq:nuf}
\end{align}
with the initial condition $ g(0)= k_{v}e(0)$.

Consider the perturbed system \eqref{eq:LparamEL}-\eqref{eq:Ybar}, with torque disturbance $p\in\mathbb{R}^{3}$, and the augmented vector of unknown parameters $\Theta = \left[ \theta^{T} , p^{T} \right]^{T}$. Then, the adaptive attitude-feedback controller is proposed as
\begin{equation}\label{eq:CtrlAd}
    \bar{\tau} = Y_{d_0} m_{0} + \bar{Y}_{d}\hat{\Theta} + k_{v}\nu - k_{p}e,
\end{equation}
where $Y_{d_{0}}\in\mathbb{R}^{4}$ and $\bar{Y}_{d}\in\mathbb{R}^{4\times 9}$ are defined in the same way as for controller \eqref{eq:AdpCtrl}. Likewise, let $\hat{\Theta}\in\mathbb{R}^{9}$ be the estimation of the augmented parameter vector, updated by the adaptive law
\begin{align}
\hat{\Theta} &= -\Gamma \bar{Y}^{T}_{d} e -\Gamma \mu \label{eq:ThetaE}\\
\dot{\mu} &=  \bar{Y}^{T}_{d} \left( e+ \nu \right) - \dot{\bar{Y}}^{T}_{d}e  \label{eq:mu}
\end{align}
with $0<\Gamma = \Gamma^{T} \in\mathbb{R}^{9\times 9}$ being an adaptation gain matrix.

Let the state be  defined as $\zeta_{3}= \left( e,\eta_{2},\nu,\tilde{\Theta},h \right) \in \mathcal{X}_{3}\vcentcolon = \mathbb{R}^{4}\times\mathbb{R}^{4}\times  \mathbb{R}^{4}\times\mathbb{R}^{9}\times\mathcal{H}$, where the discrete state $h\in\mathcal{H}$ and the parameter error $\tilde{\Theta}$ are defined in \eqref{eq:hfcn} and \eqref{eq:ThtTilD}, respectively. Then, the flow set and the jump set are given by
\begin{align}
 \bar{C}_{3} &= \{ \zeta_{3}\in\mathcal{X}_{3}   \ | \ \mathcal{G}(e) \leq \delta \}, \label{eq:SetbarC3} \\
  \bar{D}_{3} &= \{ \zeta_{3}\in\mathcal{X}_{3}   \; | \;  \mathcal{G}(e) \geq \delta \} . \label{eq:SetbarD3}
\end{align}

\begin{thm}\label{thm3}
\emph{\textbf{(Adaptive hybrid attitude-feedback controller):}}
Choose $K_{f} = K^{T}_{f} >0$, $\Gamma = \Gamma^{T}>0$, $k_{p}>0$  and
\begin{align}\label{eq:kvCond}
    k_{v} &> \frac{1}{\underbar{m}}\left( \beta + \alpha_{1} \right),
\end{align}
where
\begin{align}
    \beta \vcentcolon &= \max \left\{ \frac{\alpha^{2}_{2}}{4k_{p}} , \; \frac{\alpha^{2}_{3}k_{p} + \alpha^{2}_{2}\lambda_{\min}(K_{f})}{4k_{p}\lambda_{\min}(K_{f})} \right\} \label{eq:beta}\\
    \alpha_{1} &= k_{h_1} + 4k_{c_1} + \bar{m}\label{eq:alfa1} \\
    \alpha_{2} &= k_{h_1} + k_{h_2} + k_{c_1}\|\dot{q}_{d}\| + 4k_{c_1} + \bar{m}|k_{p}-1| \label{eq:alfa2} \\
    \alpha_{3} &= k_{h_1} + k_{c_1}\|\dot{q}_{d}\| + 4k_{c_1} + \bar{m}\|K_{f} + I_{4}\| \label{eq:alfa3}
\end{align}
and $k_{c_1}$, $k_{c_2}$, $k_{h_1}$, $k_{h_2}$, $\underbar{m}$ and $\bar{m}$ are defined in Lemma \ref{lem2} and Lemma  \ref{lem3}. Then, the control law \eqref{eq:CtrlAd} in closed loop with the system \eqref{eq:EL-modelD} renders the set $\mathcal{A}_{3}\vcentcolon = \{ (e,\eta_{2},\nu , h) \in \mathbb{R}^{4}\times\mathbb{R}^{4}\times\mathbb{R}^{4}\times \mathcal{H} | e=0_{4\times 1}, \eta_{2} = 0_{4\times 1} , \nu = 0_{4\times 1}  \}$ globally asymptotically stable, while maintaining the parameter estimation error $\tilde{\Theta}$ bounded.
\end{thm}
\begin{pf}
For all $\zeta_{3}\in\bar{C}_{3}$, the time derivative of \eqref{eq:nu}, by substituting $\dot e_f$,  %\eqref{eq:ef}, 
is given by 
\begin{align}\label{eq:nup}
    \dot{\nu} &= \mathrm{Sech}^{2}(e_{f})\dot{e}_{f} 
    = -K_{f}\nu - k_{v}\eta_{2} + k_{p}e
\end{align}
where $\mathrm{Sech}(x) \vcentcolon = \mathrm{diag}\lbrace \mathrm{sech}(x_1),\ldots ,\mathrm{sech}(x_4) \rbrace $ $\in$ $\mathbb{R}^{4\times 4}$ $\forall x\in\mathbb{R}^{4}$.

Taking the time derivative of \eqref{eq:Eta2} and premultiplying it by the inertial matrix $D(q)$ gives 
\begin{align*}
    D(q)\dot{\eta}_{2} &= D(q)\left( \ddot{e} + \dot{e} + \dot{\nu} \right) \\
    &= D(q)\ddot{q} - D(q)\ddot{q}_{d} + D(q)\dot{e} + D(q)\dot{\nu}, 
\end{align*}
where, with a little  abuse of notation,  $q_{d}$, $\dot{q}_{d}$, $\ddot{q}_{d}$  are considered instead of $hq_{d}$, $h\dot{q}_{d}$, $h\ddot{q}_{d}$ for  $\zeta_{3}\in\bar{C}_{3}$.

Substituting the system \eqref{eq:EL-modelD} and the dynamics \eqref{eq:nup} gets
\begin{align*}
    D(q)\dot{\eta}_{2} &= -C(q,\dot{q})\dot{q} + \bar{\tau} + d - D(q)\ddot{q}_{d}
    +D(q)\dot{e} \\
    &\quad + D(q)\left( -K_{f}\nu - k_{v}\eta_{2} + k_{p}e \right),
\end{align*}
which in closed loop with the control law \eqref{eq:CtrlAd} results in
\begin{equation}\label{eq:eta2P}
    D(q)\dot{\eta}_{2}  = -C(q,\dot{q})\eta_{2} + k_{v}\nu - k_{p}e -k_{v}D(q)\eta_{2} +\bar{Y}_{d}\tilde{\Theta} + \chi_{2},
\end{equation}
where 
\begin{equation}\label{eq:chi2}
    \chi_{2} = \underbar{h}(t,e,\dot{e}) + C(q,\dot{q})(e+\nu) + D(q)\left( \dot{e} - K_{f}\nu + k_{p}e\right)
\end{equation}
with $\underbar{h}(t,e,\dot{e})$  defined in \eqref{eq:hres}. 

By  Lemma \ref{lem3} and the upper bound \eqref{eq:nuUp}, it can be shown  that  $\chi_{2}$ is bounded by
\begin{equation}\label{eq:chi2UP}
    \|\chi_{2}\|\leq \alpha_{1}\|\eta_{2}\| + \alpha_{2}\|e\| + \alpha_{3} \|\nu\|,
\end{equation}
where the constants $\alpha_{1}$, $\alpha_{2}$ and $\alpha_{3}$ are defined in \eqref{eq:alfa1}-\eqref{eq:alfa3}. 

Substituting the time derivative of the parameter estimate  \eqref{eq:ThetaE} in the time evolution of the parameter  error $\tilde{\Theta} = \hat{\Theta} - \Theta$, gives 
\begin{equation}\label{eq:ThtP2}
    \dot{\Tilde{\Theta}} = \dot{\hat{\Theta}} - \dot{\Theta} = -\Gamma \bar{Y}^{T}_{d}\eta_{2}.
\end{equation}
Then the closed-loop system is given by
\begin{equation}\label{eq:ClsdLAdp2}
    \underbrace{
    \begin{array}{cl}
        \dot{e}= &  \eta_{2} - e - \nu ,\\
        D(q)\dot{\eta}_{2} = & -C(q,\dot{q})\eta_{2} + k_{v}\nu - k_{p}e -k_{v}D(q)\eta_{2} +\bar{Y}_{d}\tilde{\Theta}  \\
         & + \chi_{2},\\
        \dot{\nu} = & -K_{f}\nu - k_{v}\eta_{2} + k_{p}e,\\
        \dot{\Tilde{\Theta}} =& -\Gamma \bar{Y}^{T}_{d}\eta_{2},\\
        \dot{h} = & 0,
    \end{array}}_{\forall \zeta_{3}\in\bar{C}_{3}}
\end{equation}
\begin{equation}\label{eq:eq:ClsdLAdp2+}
    \zeta^{+}_{3}=\left[ \begin{array}{c}
         e \\
         \eta_{2} \\
         \nu \\
         \tilde{\Theta}\\
         h^{+}
    \end{array}\right] , \quad\forall \zeta_{3}\in\bar{D}_{3}.
\end{equation}
Notice that, in view of the definition \eqref{eq:nu}, the state $\nu$ does not change over jumps.

Define the following Lyapunov function candidate
\begin{equation}\label{eq:V4}
    V_{4} = \frac{1}{2}\eta^{T}_{2}D(q)\eta_{2} +\frac{k_{p}}{2}\mathcal{U}(e,h) + \frac{1}{2}\nu^{T}\nu + \frac{1}{2}\tilde{\Theta}^{T}\Gamma^{-1}\tilde{\Theta},
\end{equation}
which is radially unbounded. Its time evolution along the closed-loop system \eqref{eq:ClsdLAdp2}, for all $\zeta_{3}\in\bar{C}_{3}$, is 
\begin{align}\label{eq:V4p}
    \dot{V}_{4} &= \eta^{T}_{2}D(q)\dot{\eta}_{2} + \frac{1}{2}\eta^{T}_{2}\dot{D}(q)\eta_{2} + k_{p}e^{T}\dot{e} + \nu^{T}\dot{\nu} \notag\\
    &\quad + \tilde{\Theta}^{T}\Gamma^{-1}\dot{\tilde{\Theta}} \notag\\
    &= \eta^{T}_{2}\left( -C(q,\dot{q})\eta_{2} + k_{v}\nu - k_{p}e -k_{v}D(q)\eta_{2} +\bar{Y}_{d}\tilde{\Theta}\right.  \notag \\
    &\quad + \chi_{2}  \big) + \frac{1}{2}\eta^{T}\dot{D}(q)\eta + k_{p}e^{T}\left( \eta_{2}-e-\nu \right) \notag\\
    &\quad + \nu^{T}\left( -K_{f}\nu - k_{v}\eta_{2} + k_{p}e \right) +\tilde{\Theta}\Gamma^{-1}\left( -\Gamma \bar{Y}^{T}_{d}\eta_{2}\right) \notag\\
    &= -k_{v}\eta^{T}_{2}D(q)\eta_{2} + \eta^{T}_{2}\chi_{2} - k_{p}e^{T}e -\nu^{T}K_{f}\nu  \notag \\
    &\leq -k_{v}\underbar{m}\|\eta_{2}\|^{2} + \|\eta_{2}\|\|\chi_{2}\| -k_{p}\|e\|^{2} \notag\\
    &\quad -\lambda_{\min}(K_{f})\|\nu\|^{2} \notag \\
    &\leq -k_{v}\underbar{m}\|\eta_{2}\|^{2} + \|\eta_{2}\|\left(  \alpha_{1}\|\eta_{2}\| + \alpha_{2}\|e\| + \alpha_{3} \|\nu\|\right) \notag\\
    &\quad -k_{p}\|e\|^{2}  -\lambda_{\min}(K_{f})\|\nu\|^{2} \notag \\
    &\leq - z^{T} \Phi_{2} z,
\end{align}
where \eqref{pLg2} of Lemma \ref{lem2} and \eqref{eq:chi2UP} are  used,  $z: = \left[ \|\eta_{2} \| ,\; \|e \|, \; \|\nu \| \right]^{T} $, and
\begin{equation}\label{eq:FI}
    \Phi_{2} \vcentcolon = \left[
    \begin{array}{ccc}
         k_{v}\underbar{m} - \alpha_{1}& -\frac{\alpha_{2}}{2} & -\frac{\alpha_{3}}{2} \\
         -\frac{\alpha_{2}}{2} & k_{p} & 0 \\
         -\frac{\alpha_{3}}{2} & 0 & \lambda_{\min}(K_{f})
    \end{array}
    \right].
\end{equation}

Under condition \eqref{eq:kvCond},   matrix $\Phi_2$ is positive definite. Therefore, the time derivative \eqref{eq:V4p} is negative definite. This proves that the states $\zeta_{3}\in\bar{C}_{3}$ are bounded.

In the jump set $\zeta_{3}\in\bar{D}_{3}$, in view of \eqref{eq:eq:ClsdLAdp2+}, it has
\begin{align}
    V_{4}\left(\zeta^{+}_{3}\right) - V_{4}(\zeta_{3}) &=   \frac{k_{p}}{2}\mathcal{U}(e,h^{+}) - \frac{k_{p}}{2}\mathcal{U}(e,h) \notag\\
    &\leq  -\frac{k_{p}}{2}\delta . \label{eq:diffonDAdp}
\end{align}

In consequence, $V_{4}(\zeta_{3})$ is  non-increasing  for all $\zeta_{3}\in\bar{C}_{3}$, and strictly decreasing over the set $\bar{D}_{3}$. Therefore, under the same arguments of the proof of Theorem \ref{thm2}, it concludes  the global asymptotic convergence of $\dot{V}_{4}\to 0$.
\begin{flushright}
$\square$
\end{flushright}
\end{pf}

\begin{rem}
\emph{\textbf{(Adaptive hybrid attitude-feedback controller):}}
The proposed 4-DOF Lagrangian dynamics facilitates achieving the global asymptotic stability of the equilibrium $e = 0_{4\times 1}$, $\dot{e} = 0_{4\times 1}$ in  overall system under the adaptive attitude-feedback control law \eqref{eq:CtrlAd}, and  results in a relative simpler design  as compared to other adaptive attitude  controllers 
\cite{slotine1990hamiltonian,wen1991attitude,egeland1994passivity,ahmed1998adaptive,wong2001adaptive,costic2001quaternion,luo2005inverse}. Moreover, by leveraging the nonlinear filter \eqref{eq:nu},  the requirement of the angular velocity measurements is not needed in the proposed adaptive controller.
\end{rem}

\section{Simulations}\label{Sec:Sim}

Two simulations were carried out. The first simulation is aimed at illustrating the unwinding phenomenon  of the state-feedback control \eqref{eq:Ctrl} and  lack of robustness of the discontinuous controller (controller \eqref{eq:CtrlHyb} with the gap exceeding $\delta=0$)  when the quaternion measurement is noisy, and how these facts are eliminated by the hybrid state-feedback controller \eqref{eq:CtrlHyb} by a proper choice of $\delta$.  The second simulation is to show the performance of the adaptive hybrid attitude-feedback controller \eqref{eq:CtrlAd} for two values of the gap exceeding $\delta$ under situations of noise-free and noisy attitude measurements. In both simulations, the torque input applied to the spacecraft is obtained by \eqref{eq:Tau}.

\subsection{Simulation 1: state-feedback controllers}
    The desired trajectory was  given  by   \eqref{eq:qd} with $\omega_{d}=0_{3\times 1} \ [\mathrm{rad}/s]$ $\forall t\geq 0 \ [s]$ and the initial condition  $q_{d}^T(0) = [1,0_{1\times 3}]$.  
    The inertial matrix in \eqref{eq:M0} was $M=\mathrm{diag}\lbrace 10 \bar{u} \rbrace$ [$\mathrm{Kg}\;\mathrm{m}^{2}$],  $m_0=1$, where $\bar{u}=u/\| u \|$ and $u^T=[1,2,3]$.
    The controller gains \eqref{eq:Ctrl} and \eqref{eq:CtrlHyb} were set as  $\Lambda = 0.1 I_{4}$,  $K_{s} = I_{4}$.
The quaternion measurement was given by  $q_{m}=(q+n \bar{v})/\|q+n \bar{v} \|$, $\bar{v}=v/ \| v \|$, where $v \in \mathbb{R}^{4}$  is a zero-mean Gauss distribution with variance $0.2$, and $n$ an uniform distribution. Two scenarios were simulated with the settings shown in Table \ref{tab:Inertia}. 

\begin{table}[hbt!]
    \caption{Scenario settings for the first simulation.}
    \centering
	\begin{tabular}{clllll}\hline
	Scenario	 & $q(0)$                & $\omega(0)$ & $\epsilon_0(0)$ &  $h(0)$ & $q_m$                    \\ \hline
	    
	  1.1          & $[0, \bar{u}^{T}]^{T}$ & $0_{3\times 1}$& $0$             &  $1$      &   $n=0$     \\  
	  1.2          & $[0, \bar{u}^{T}]^{T}$ & $0.5\bar u$    & $0$             & $1$      &   $n \in [0,  0.1]$        \\
	  \hline
	\end{tabular}
	\label{tab:Inertia}
\end{table}

\begin{figure}[tbh!]
	\begin{center}
		\includegraphics[trim = 10mm 6mm 0mm 0mm,scale=0.28]{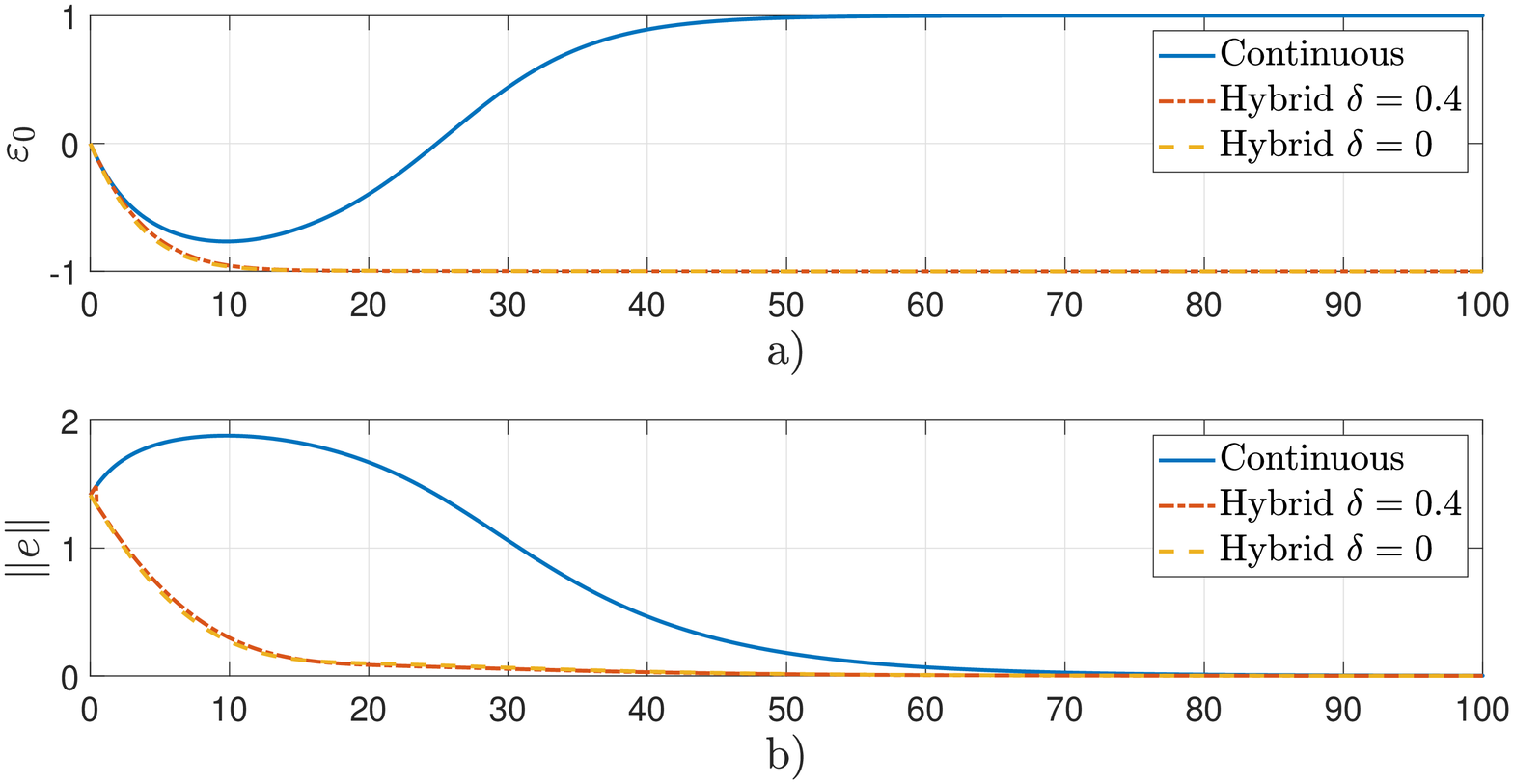}
		\includegraphics[trim = 10mm 6mm 0mm 0mm,scale=0.28]{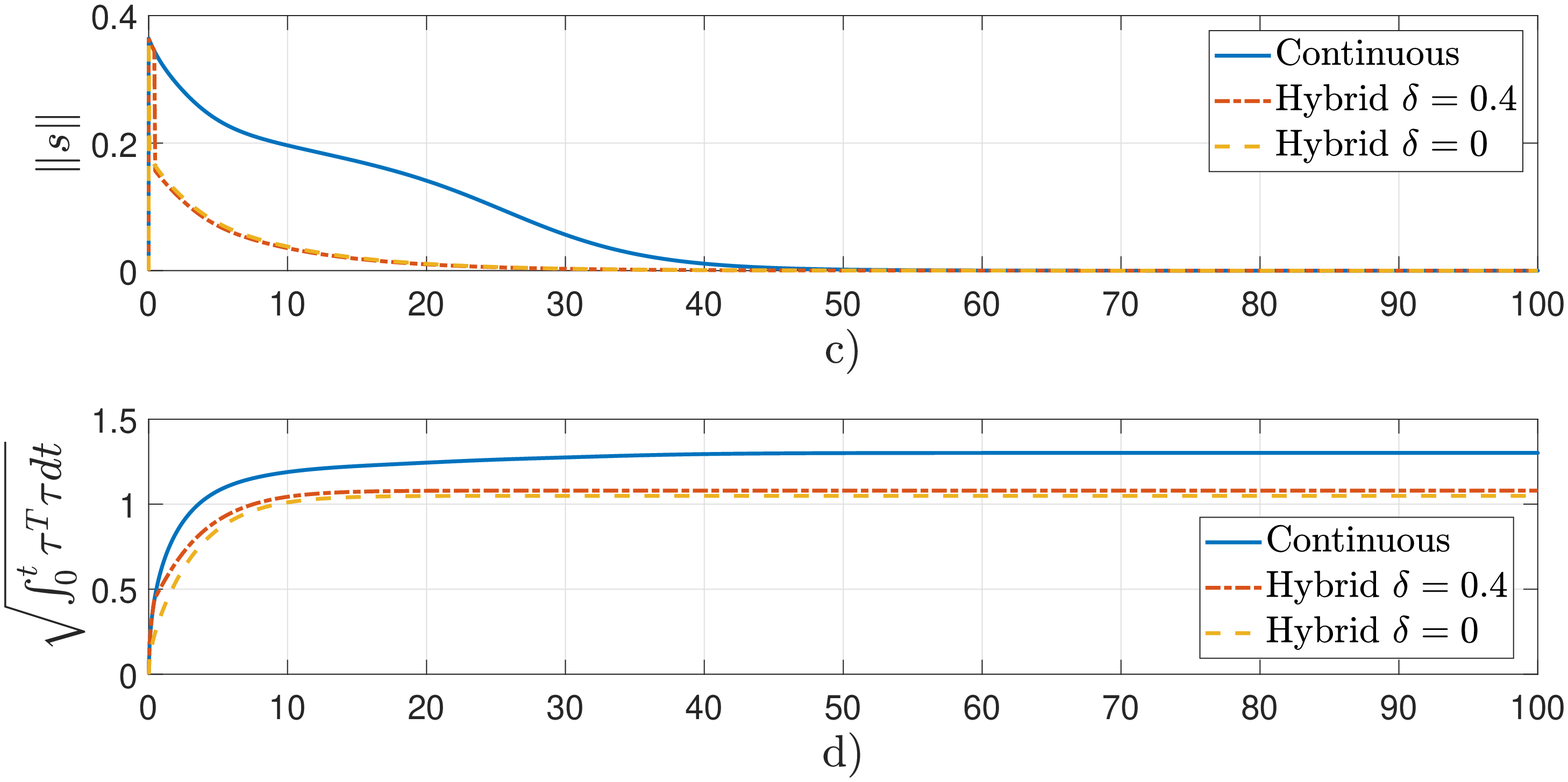}
		\includegraphics[trim = 10mm 80mm 0mm 0mm,scale=0.28]{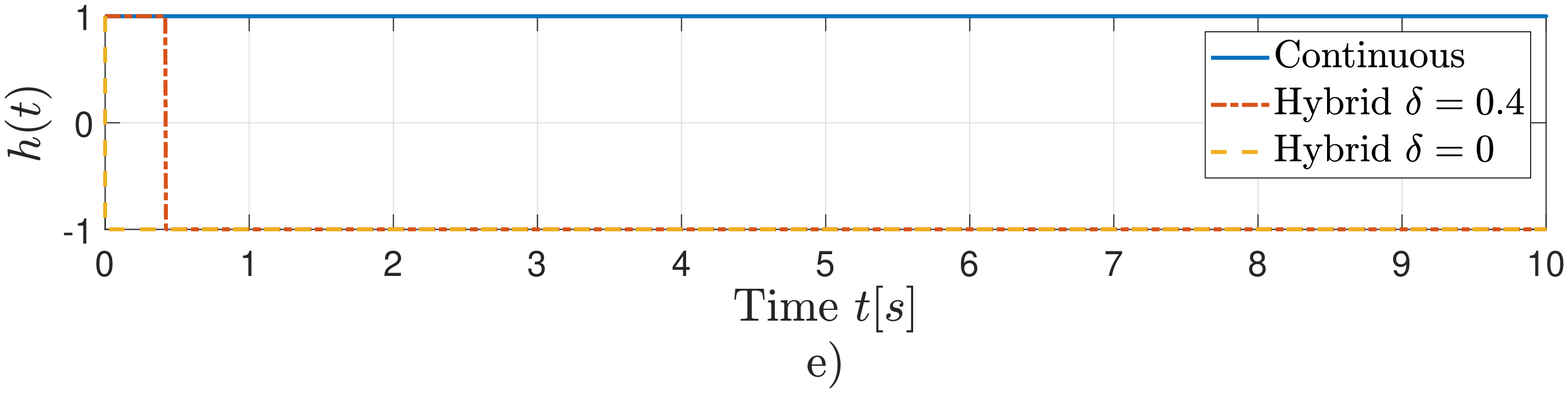}
		\caption{Scenario 1.1: Performance of state-feedback  controllers \eqref{eq:Ctrl},  the discontinuous controller \eqref{eq:CtrlHyb} with the gap exceeding $\delta = 0$, and the hybrid controller \eqref{eq:CtrlHyb} with $\delta = 0.4$ and. Unwinding phenomenon is observed in the continuous state-feedback controller.}
		\label{fig:Sc1Unw}
	\end{center}
\end{figure}

In Scenario 1.1  the quaternion measurement is noise-free, i.e., $q_m=q$, and the initial angular velocity $\omega(0)=0.5\bar u$, which favored to stabilize  $\varepsilon_{0}\to -1$ as observed in Fig. \ref{fig:Sc1Unw}(a). This forced  the continuous controller \eqref{eq:Ctrl} to hold  the attitude at the beginning, and then to push it to $\varepsilon_{0} = 1$. Since $h$ is kept  a constant for the continuous controller \eqref{eq:Ctrl}, no jumps were made to stabilize $\varepsilon_{0} = -1$, causing a full rotation known as unwinding phenomenon (Fig. \ref{fig:Sc1Unw}(a)). This phenomenon  was removed by the hybrid controller \eqref{eq:CtrlHyb} by performing  a jump of the discrete state $h$ at $0$ $[s]$ when the gap exceeding $\delta = 0$. Notice that this case corresponds to the discontinuous controller of \cite{mayhew2011quaternion}, and is known  not robust to small noise in the measurements.  After changing the gap exceeding  to $\delta = 0.4$,  the hybrid controller \eqref{eq:CtrlHyb} made a jump of the discrete state $h$ at $5\ [s]$  (Fig. \ref{fig:Sc1Unw}(e)), stabilizing $\varepsilon_{0} = -1$, (Fig. \ref{fig:Sc1Unw}(a)). Note that the exponential convergence of the equilibrium $(e,s)=(0_{4\times 1},0_{4\times 1})$ is  achieved at $30$ $[s]$ for the hybrid controllers, compared with   $60$ $[s]$  for the continuous controller (Fig. \ref{fig:Sc1Unw}(b) and (c)). Moreover, the continuous controller has a $23\%$ more in energy consumption  measured by  $\sqrt{\int^{t}_{0}\tau^{T}\tau d\uptau }$.

%\subsubsection{Noisy-measurement case} % with zero initial angular velocity}

\begin{figure}[tbh!]
	\begin{center}
		\includegraphics[trim = 10mm 6mm 0mm 0mm,scale=0.28]{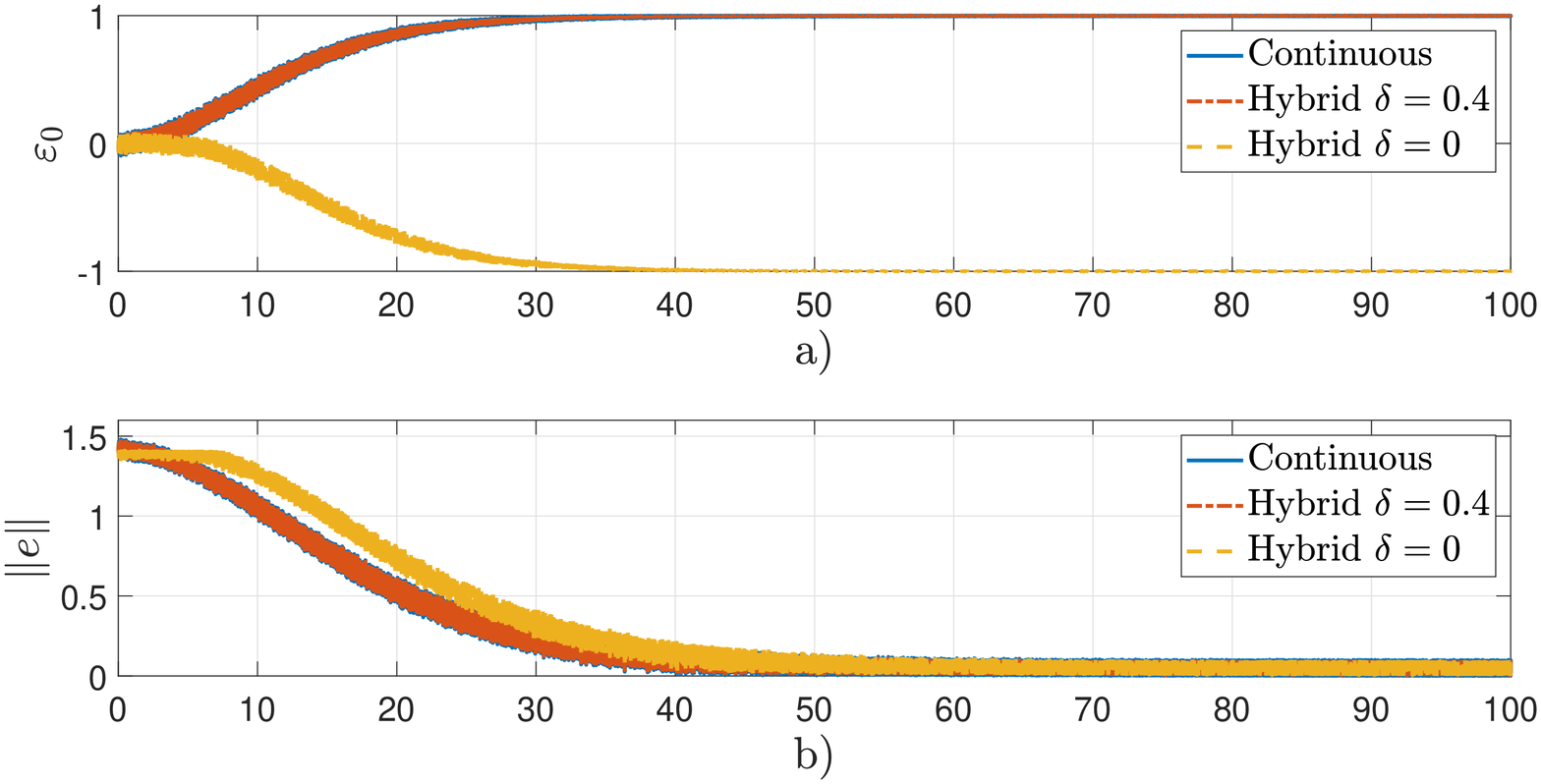}
		\includegraphics[trim = 10mm 6mm 0mm 0mm,scale=0.28]{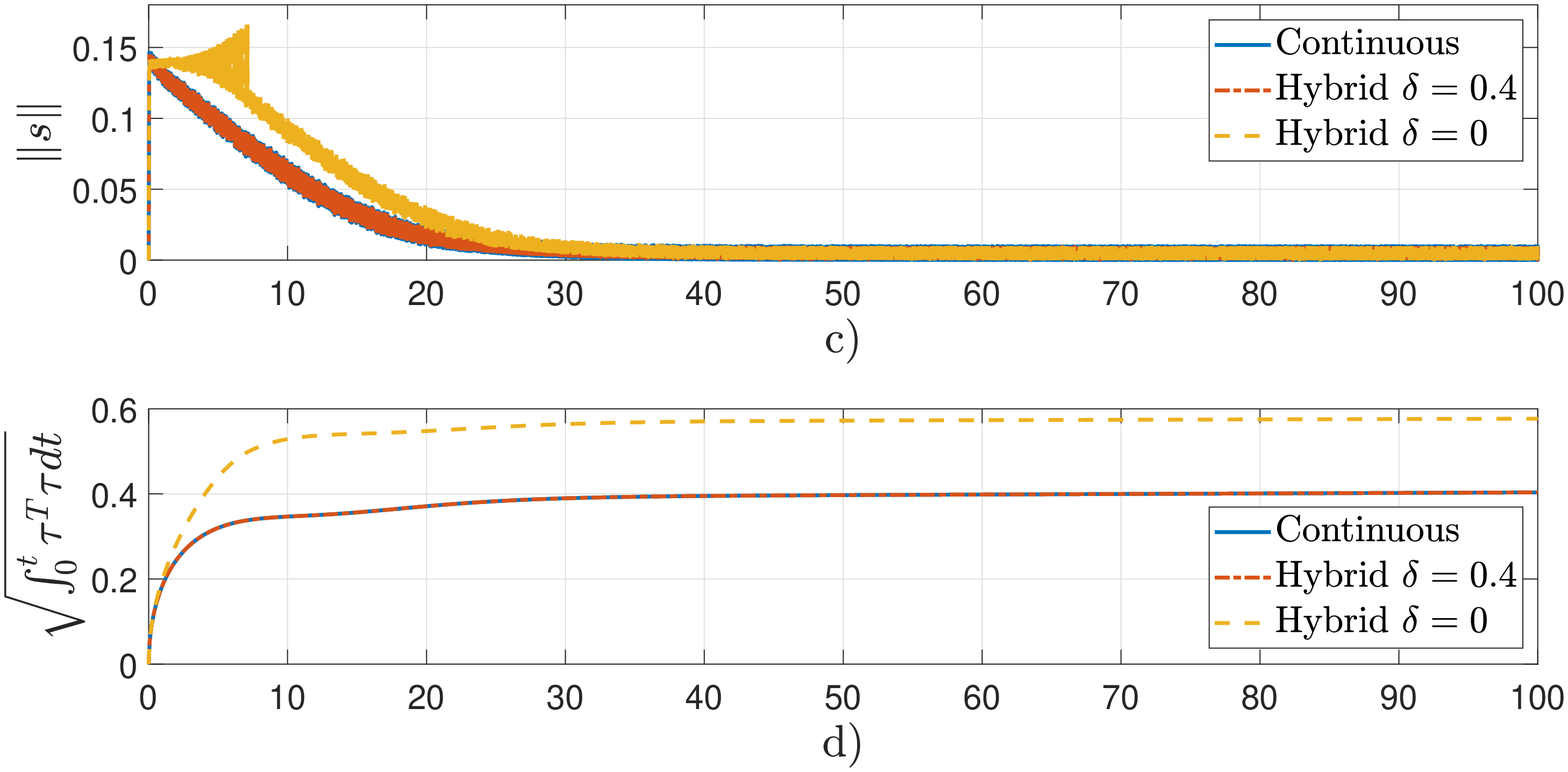}
		\includegraphics[trim = 10mm 80mm 0mm 0mm,scale=0.28]{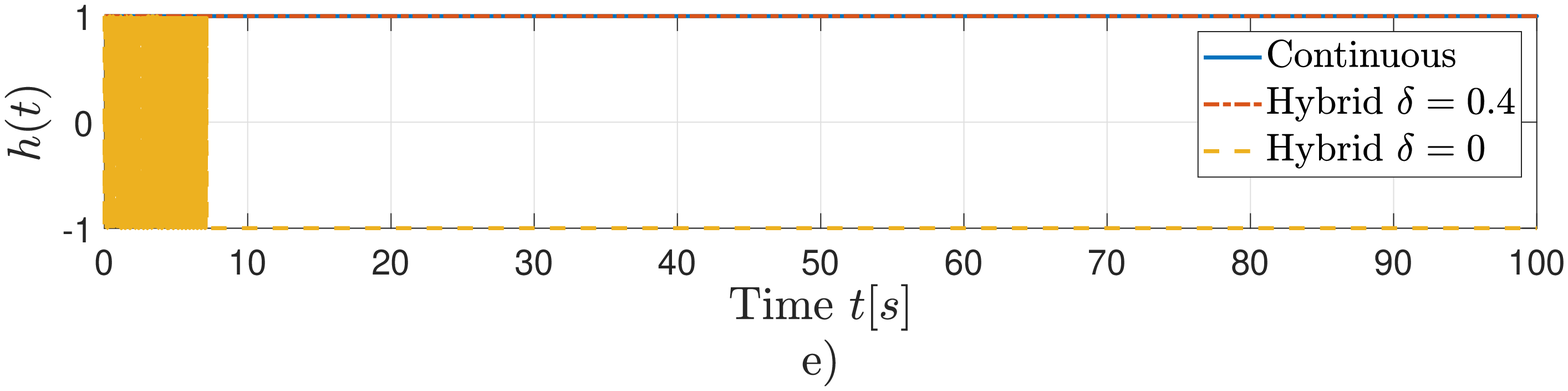}
		\caption{Scenario 1.2: Performance of the continuous state-feedback   controllers \eqref{eq:Ctrl},  the discontinuous controller \eqref{eq:CtrlHyb} with  $\delta = 0$, and the hybrid controller \eqref{eq:CtrlHyb} with $\delta = 0.4$ and respectively. The lack of robustness to noisy measurements in the discontinuous controller is shown.}
		\label{fig:Sc1Nys}
	\end{center}
\end{figure}

The lack of robustness of the discontinuous controller \eqref{eq:CtrlHyb} when the gap exceeding $\delta=0$ in the presence of noise in the quaternion measurement is  illustrated in Scenario 1.2 of Table \ref{tab:Inertia}. Observe  that the presence of noise  affects severely the hybrid controller when  $\delta = 0$, where it shows a  chattering phenomenon for the first $9$ $[s]$ reflected in the discrete state $h(t)$ (Fig. \ref{fig:Sc1Nys}(e)), causing a delay in the error convergence (Fig. \ref{fig:Sc1Nys}(a),(b),(c)) and more energy consumption  (Fig. \ref{fig:Sc1Nys}(d)). In fact, the hybrid controller with the gap exceeding  $\delta = 0$ had an increase of $45\%$ in energy consumption than the other controllers.  The sensitivity to noise was eliminated when the gap exceeding was changed to $\delta = 0.4$. Note that  the discrete state makes no  jumps despite of the noisy measurements, keeping  the discrete state $h=1$ $\forall t\geq 0$ (Fig. \ref{fig:Sc1Nys}(e)). Notice that for this scenario, the hybrid controller with the gap exceeding  $\delta = 0.4$ and the continuous controller have the same behavior, stabilizing $\varepsilon_{0} = 1$ at $30$ $[s]$ (Fig.\ref{fig:Sc1Nys}(a)). The exponential convergence of the equilibrium point $(e,s)=(0_{4\times 1},0_{4\times 1})$ is shown in Fig. \ref{fig:Sc1Nys}(b) and Fig. \ref{fig:Sc1Nys}(c).

\subsection{Simulation 2: the adaptive hybrid attitude-feedback controller}
This simulation shows the behavior of the adaptive hybrid attitude-feedback  controller \eqref{eq:CtrlAd} for different values of the gap exceeding and how this behaviour is affected by noisy  quaternion measurements and time-varying torque disturbances.  The desired trajectory is generated by \eqref{eq:qd} with the initial condition $q_d^T(0)=[1,0_{1\times 3}]$ and the desired angular velocity  $\omega_d = 0.1\sin (0.2\pi t) \left[ 1,1,1 \right]^{T}$ $\left[ \mathrm{rad}/s \right]$ used in \cite{tayebi2008unit}. 
The initial conditions were $q(0)=[0,0,1,0]^T$, $\omega(0)=\bar u$ with $\bar u=u/\|u\|$ and $u=[1, 2, 3]^T$. The initial discrete state was set $h(0)=1$ for all scenarios. The controller gains \eqref{eq:CtrlAd} were $k_v = 3$ and,  $k_p=0.7$, with a filter gain \eqref{eq:pP}, $K_f=0.1I_4$, and the adaptation gain \eqref{eq:ThetaE} was $\Gamma = \mathrm{diag}\{1000I_3, I_6\}$. The unknown inertial matrix \eqref{eq:M0} and the  noisy quaternion measurement were the same as in the first simulation. In addition, a torque disturbance $\dot{p}=v$ was added (see \eqref{eq:dDist}), with the initial condition $p(0)=[0.2,-0.1,-0.05]^{T} \ [\mathrm{Nm}]$, where $v=0_{3\times 1}$ for a constant disturbance and  $v \in \mathbb{R}^{4}$ whose element is a zero-mean Gauss distribution with variance $0.2$ for a time-varying disturbance (see Table \ref{tab:InitialV}).

\begin{table}[hbt!]
    \caption{Scenario settings for the the hybrid attitude-feedback controller \eqref{eq:CtrlAd}.}
    \centering
	\begin{tabular}{cllllll}\hline
	Scenario	 & $\delta$                & $q_m$       &  $p(t)$ \\ \hline
	  2.1          & $0.9$                 &   $n=0$     &  $[0.2,-0.1,-0.05]^{T} $      \\  
	  2.2          & $0.4$                 &   $n=0$     & $[0.2,-0.1,-0.05]^{T} $      \\ 
	  2.3          & $0.4$                 &   $n\in[0,0.1]$  &  $[0.2,-0.1,-0.05]^{T} $     \\ 
	  2.4          & $0.4$                 &   $n=0$  &  $\dot{p}=v$, $v\in N(0,0.2)$     \\ 
	  \hline
	\end{tabular}
	\label{tab:InitialV}
\end{table}

\begin{figure}[tbh!]
	\begin{center}
		\includegraphics[trim = 10mm 6mm 0mm 0mm,scale=0.34]{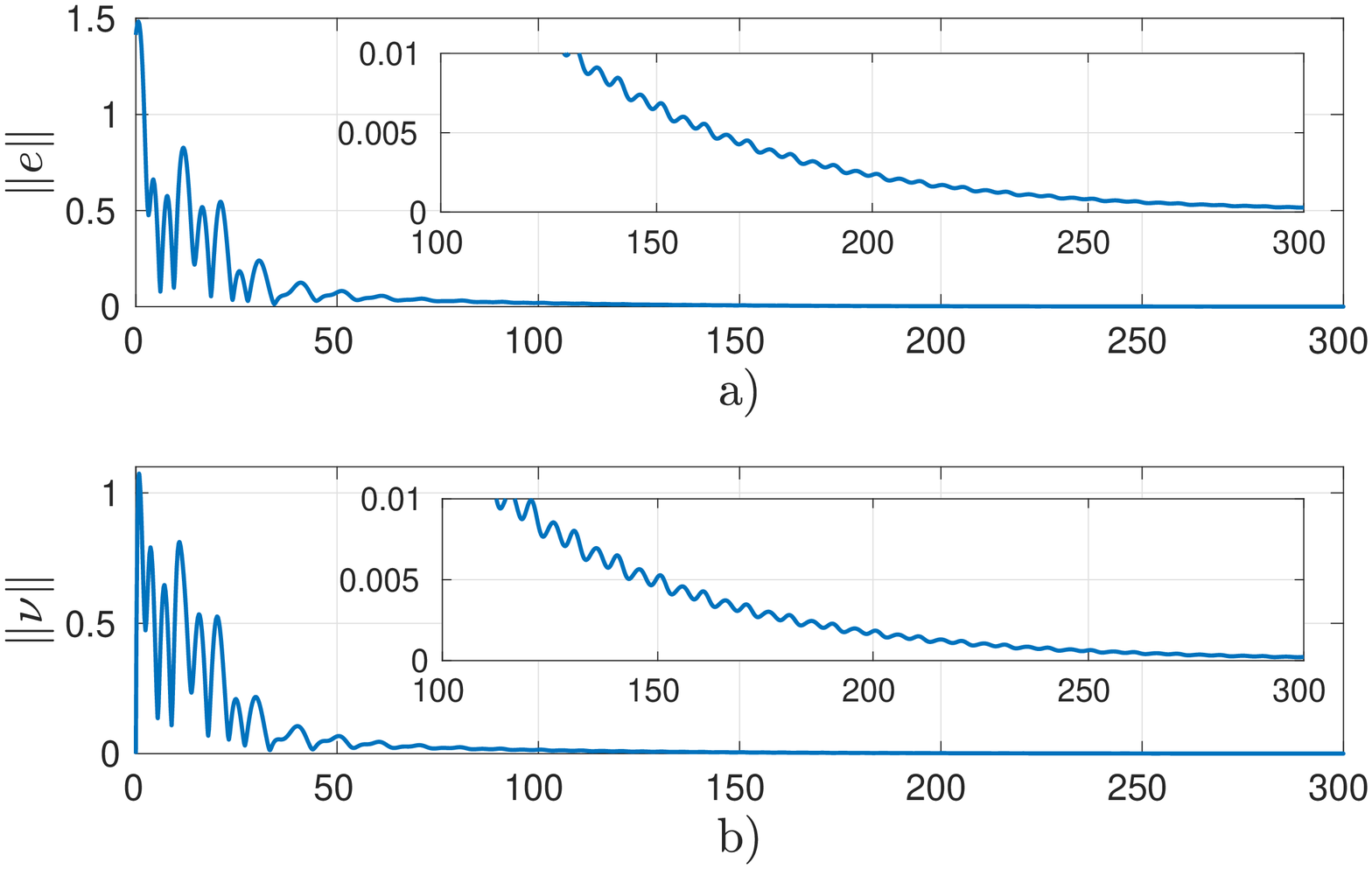}
		\includegraphics[trim = 10mm 6mm 0mm 0mm,scale=0.34]{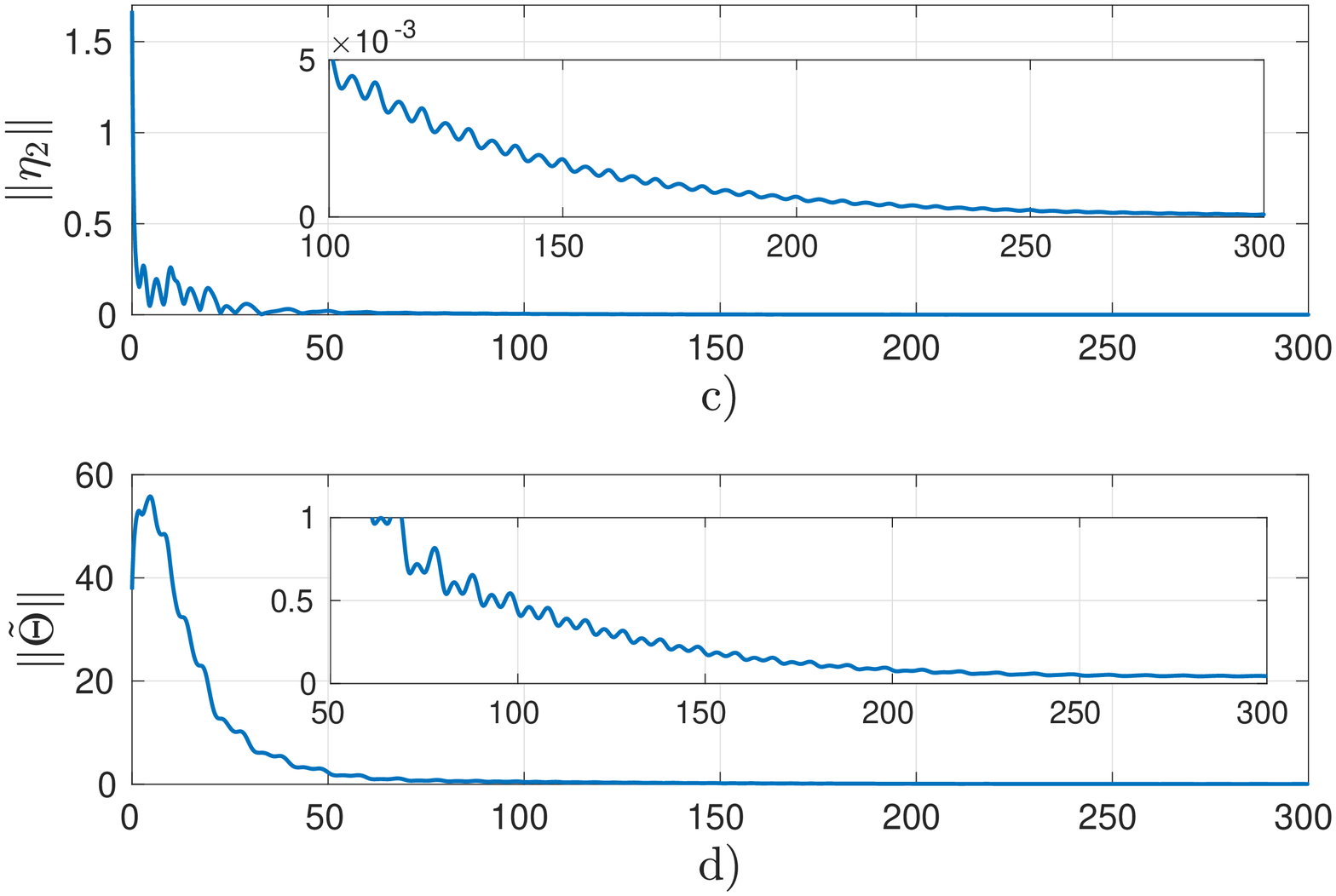}
		\includegraphics[trim = 10mm 6mm 0mm 0mm,scale=0.34]{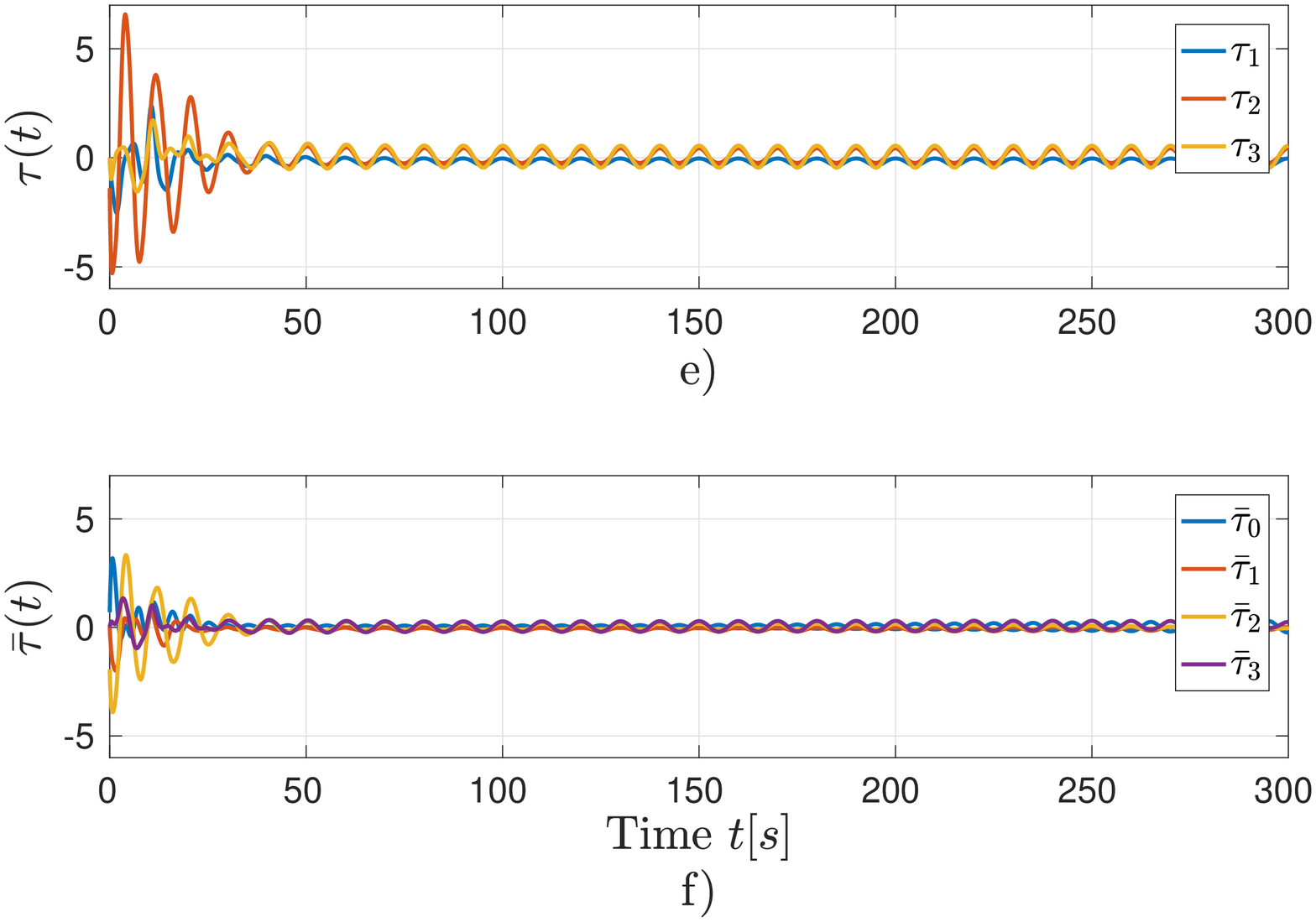}
		\caption{Scenario 2.1: Performance of the adaptive hybrid attitude-feedback controller \eqref{eq:CtrlAd} with a large gap exceeding ($\delta = 0.9$).}
		\label{fig:AdpID}
	\end{center}
\end{figure}

%\subsubsection{Noise-free situation with a large gap exceeding}
Scenario 2.1 illustrates the performance of the proposed adaptive hybrid attitude-feedback controller for the gap exceeding $\delta = 0.9$ under the noise-free measurements and constant torque disturbance.  Due to the large gap exceeding, no jumps were made in spite of the initial angular velocity $\omega(0)$ which favored to stabilizing $\epsilon_0=-1$. The norm of $\| e (t) \|$, $\|\nu(t)\|$ and $\|\eta_{2}(t)\|$ are drawn in Figs. \ref{fig:AdpID}(a), (b) and (c), respectively.  Notice that the estimation error $\|\tilde{\Theta}(t)\|$ remained  close to zero (Fig. \ref{fig:AdpID}(f)), converging to $0.04$, as ensured by Theorem \ref{thm3}. The generalized torque $\bar{\tau}(t)$ and the torque applied to the spacecraft $\bar{\tau}(t)$ are depicted in Fig. \ref{fig:AdpID}(g) and (h).

\begin{figure}[tbh!]
	\begin{center}
		\includegraphics[trim = 10mm 6mm 0mm 0mm,scale=0.34]{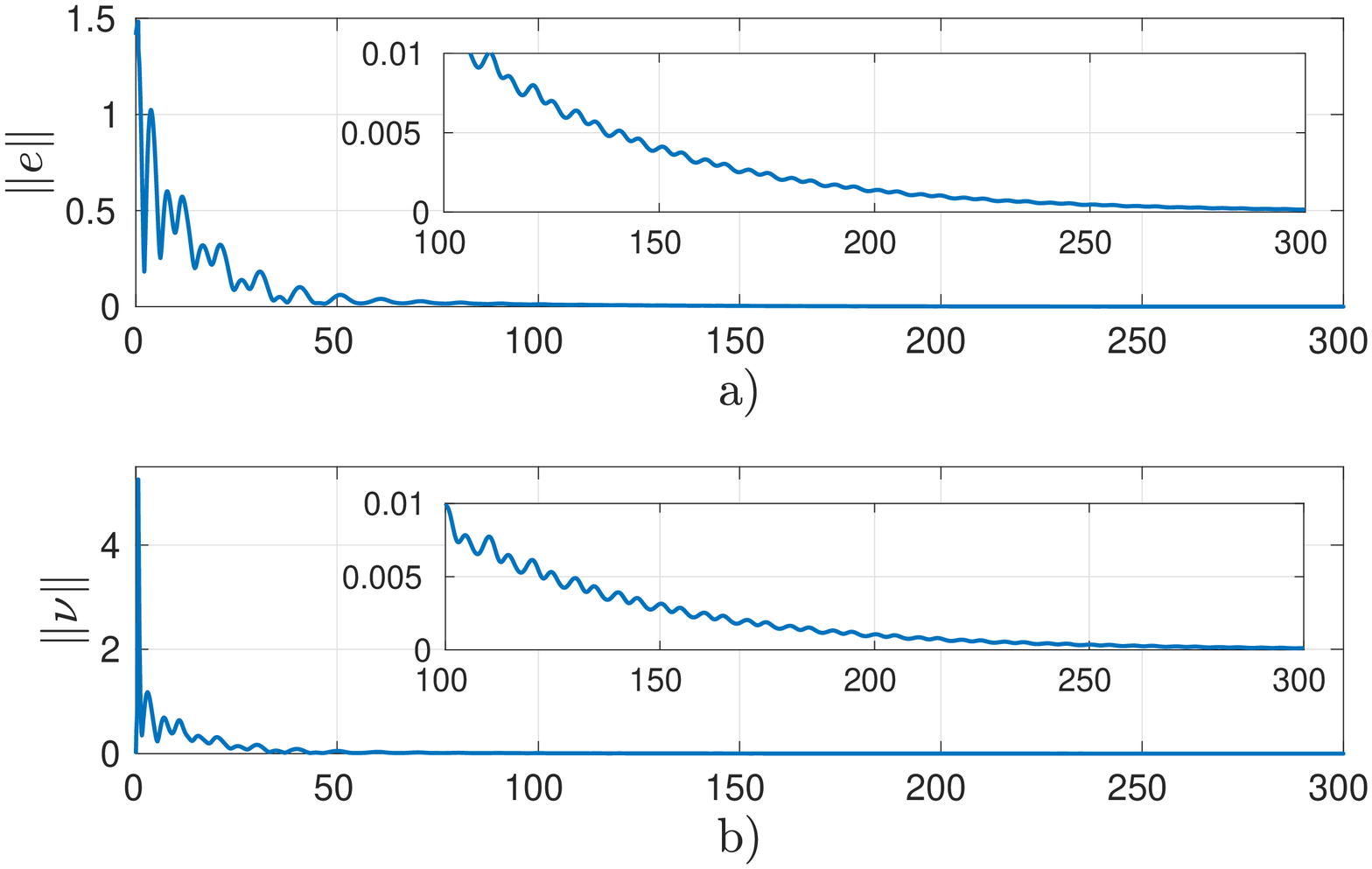}
		\includegraphics[trim = 10mm 6mm 0mm 0mm,scale=0.34]{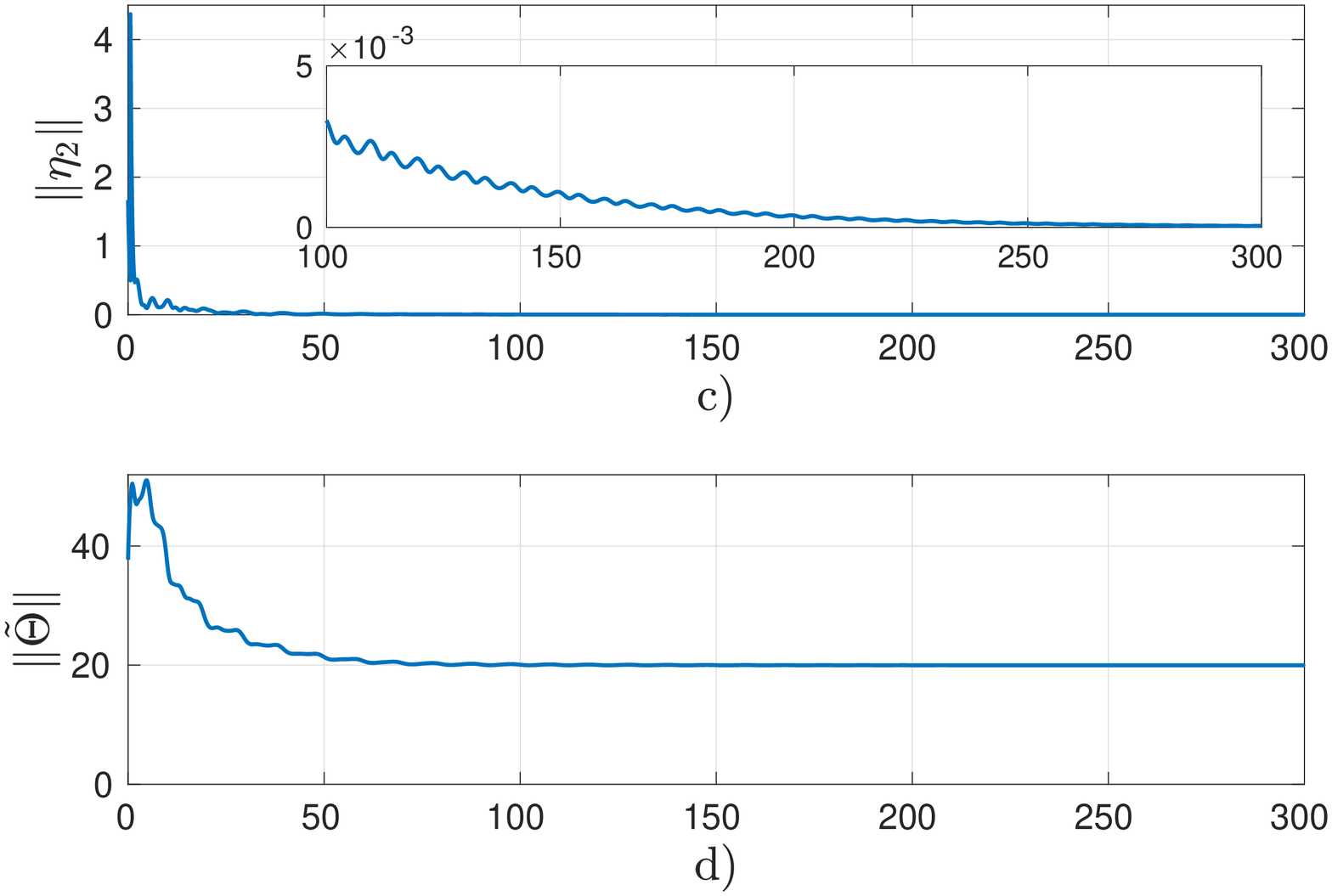}
		\includegraphics[trim = 10mm 6mm 0mm 0mm,scale=0.34]{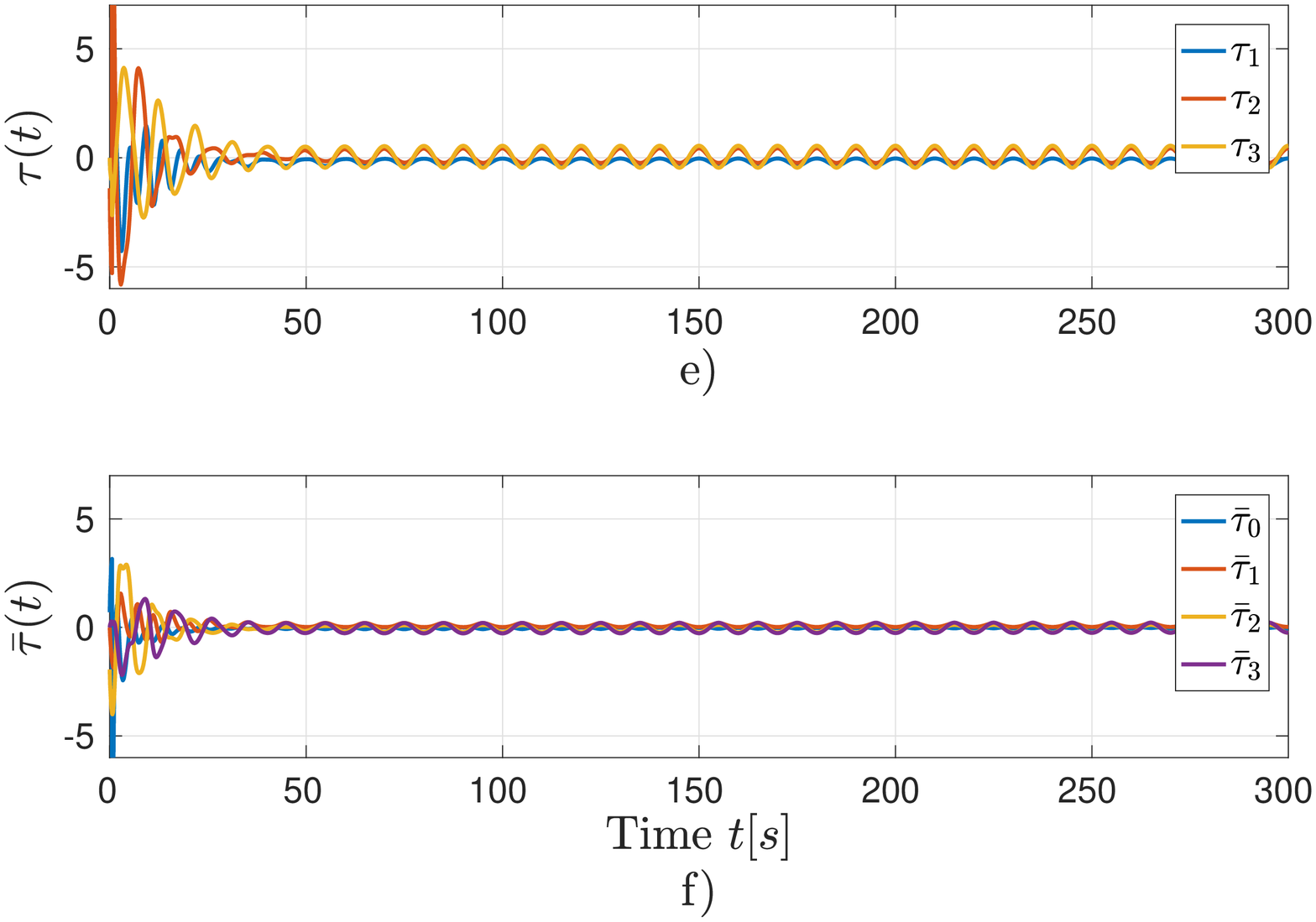}
		\caption{Scenario 2.2:  Performance of the adaptive hybrid attitude-feedback controller \eqref{eq:CtrlAd} with a small gap exceeding ($\delta = 0.4$).}
		\label{fig:AdpSW}
	\end{center}
\end{figure}

%\subsubsection{Noise-free situation with a small gap exceeding }
When the gap exceeding was reduced to $\delta=0.4$, the discrete state $h$ to perform a jump at $t=0.5\ [s]$ due to the  initial angular velocity $\omega (0)=\bar{u}$,  forcing  the attitude  to tack a closer path to stabilize $\varepsilon_{0}\to -\hat{1}$ as shown in Figs. \ref{fig:AdpSW}(a), (b) and (c), where  the norm of  $\|e\|$, $\|\nu\|$, and $\|\eta_{2}\|$ showed  a faster convergence compared to the previous scenario. However, the parameter estimation error $\tilde{\Theta}$ converged to $20$ since  the states $e$ and $\nu$ were  less exciting compared with the previous situation in the first $50$ $[s]$. Also, notice that the control torque $\tau$ and the generalized torque $\bar{\tau}$ increased slightly in the first $20$ $[s]$ (\ref{fig:AdpSW}(e), (f)).

\begin{figure}[tbh!]
	\begin{center}
		\includegraphics[trim = 10mm 6mm 0mm 0mm,scale=0.34]{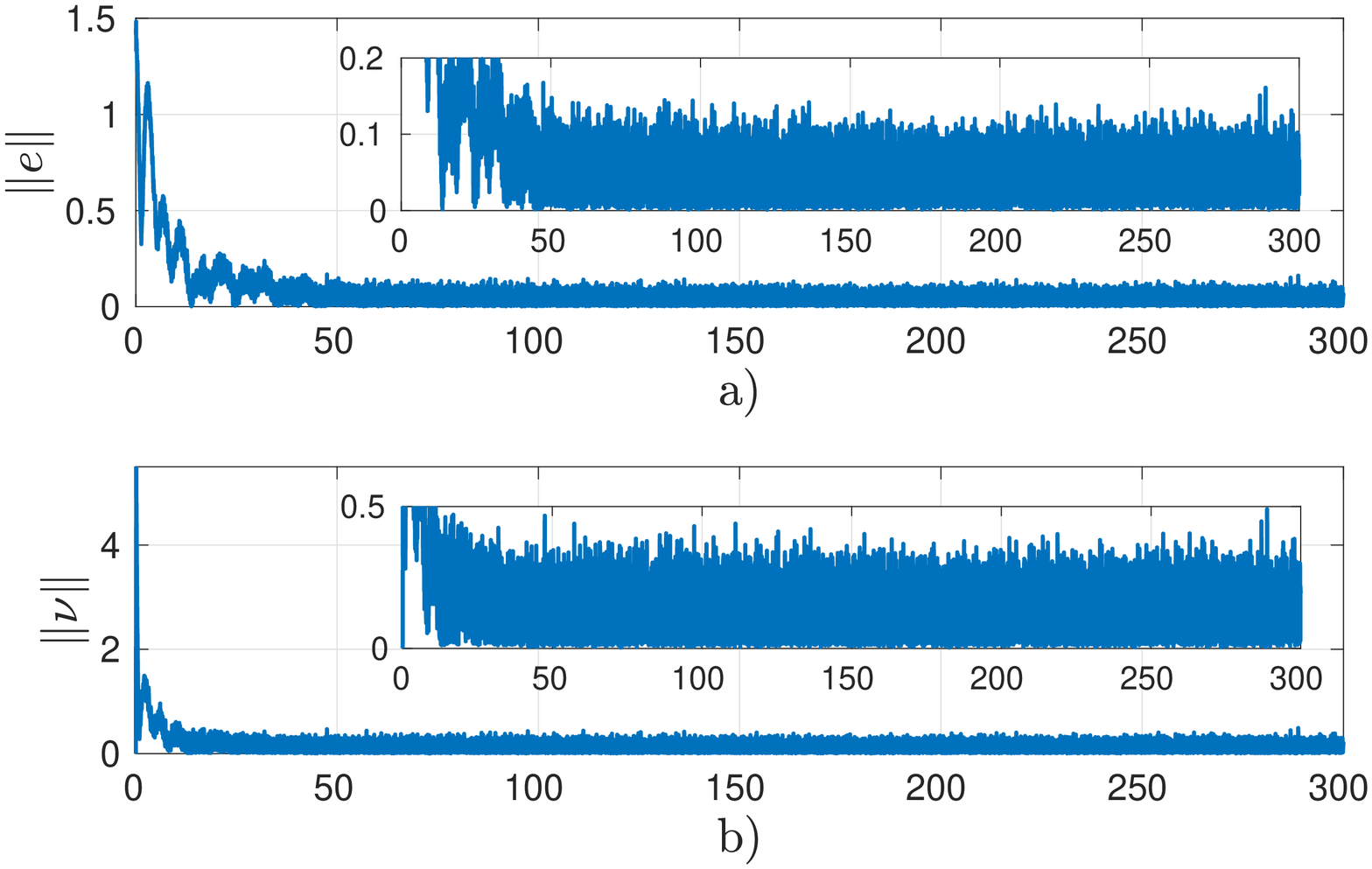}
		\includegraphics[trim = 10mm 6mm 0mm 0mm,scale=0.34]{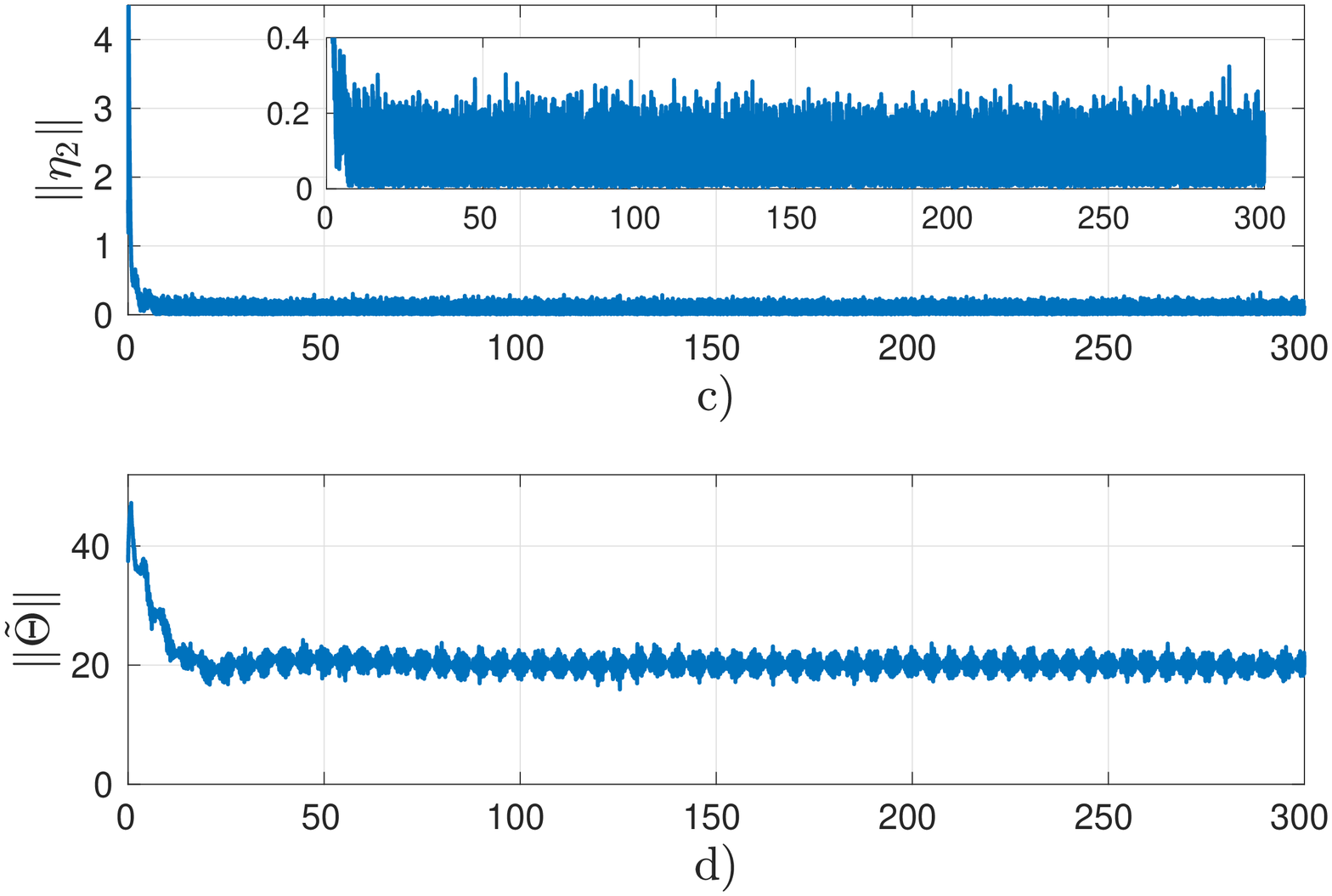}
		\includegraphics[trim = 10mm 6mm 0mm 0mm,scale=0.34]{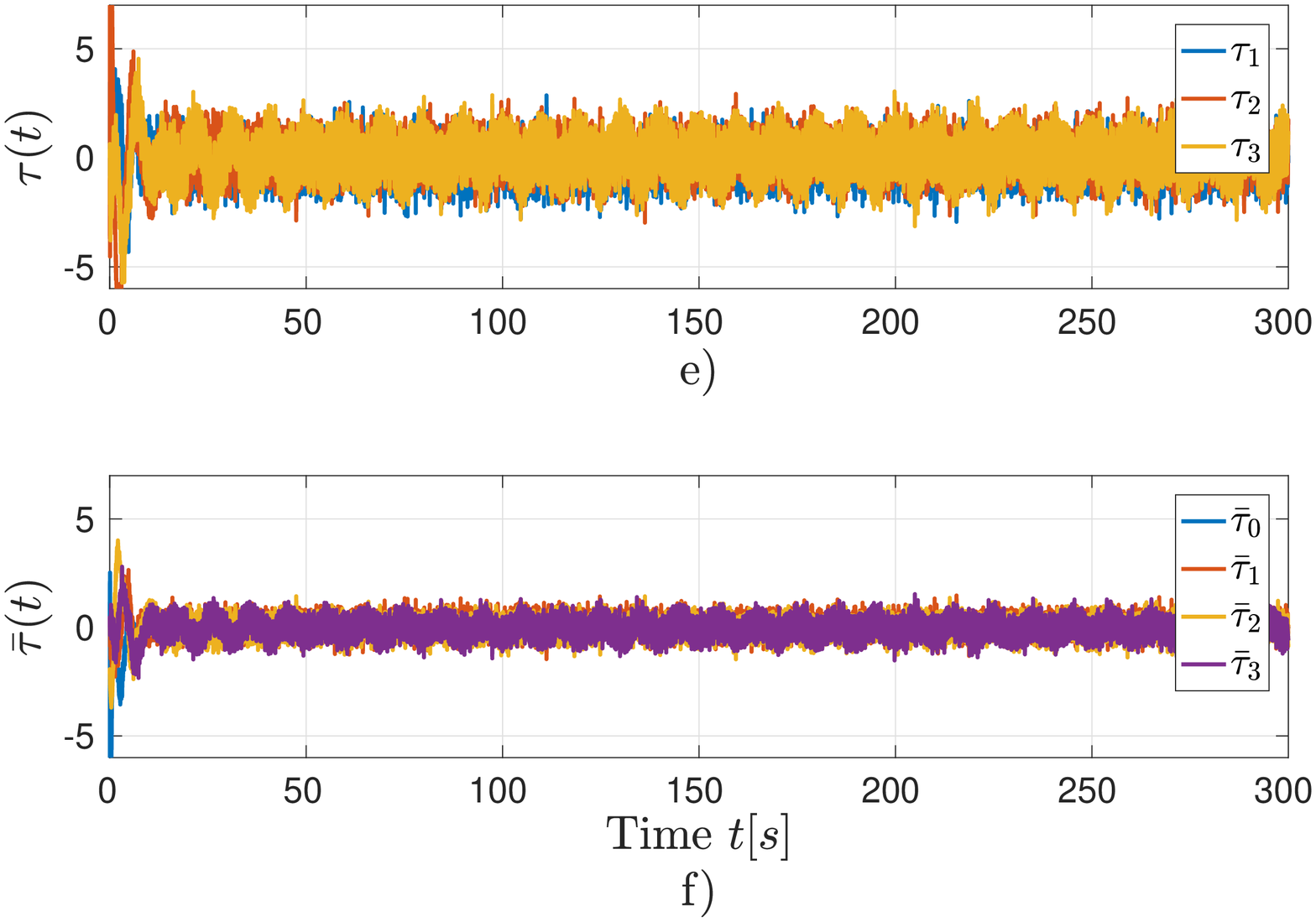}
		\caption{Scenario 2.3: Performance of adaptive controller \eqref{eq:CtrlAd} with gap exceeding $\delta = 0.4$ under noisy attitude measurements.}
		\label{fig:AdpNys}
	\end{center}
\end{figure}

%\subsubsection{Noisy measurement situation}
Scenario 2.3 shows the performance of the adaptive hybrid attitude-feedback controller \eqref{eq:CtrlAd} when  the gap exceeding was kept to $\delta=0.4$ and the noisy attitude measurement $q_{m}$ as in Scenario 1.2. Observe from Fig. \ref{fig:AdpNys}(a), (b) and (c) that the  norm of  $\|e\|$, $\|\nu\|$ and $\|\eta_{2}\|$ showed  an asymptotically decaying behavior as in the  previous scenarios,  remained bounded by $0.2$, $0.5$ and $0.4$, respectively, after $50\ [s]$. Note that the state $\nu(t)$ was most affected by the attitude noise,  because it contains a term in  \eqref{eq:nuf} that depends directly on the error $e(t)$ multiplied by the gain $k_v$, which amplifies  the noise. The estimation error $\tilde{\Theta}$ remained bounded oscillating around  $20$ (Fig. \ref{fig:AdpNys}(d)).   Figs.  \ref{fig:AdpNys}(e) and (d) draw the control torques, which  increased about $1.0$ $[\textrm{Nm}]$ respect to the previous  noise-free scenarios.  

\begin{figure}[tbh!]
	\begin{center}
		\includegraphics[trim = 10mm 6mm 0mm 0mm,scale=0.34]{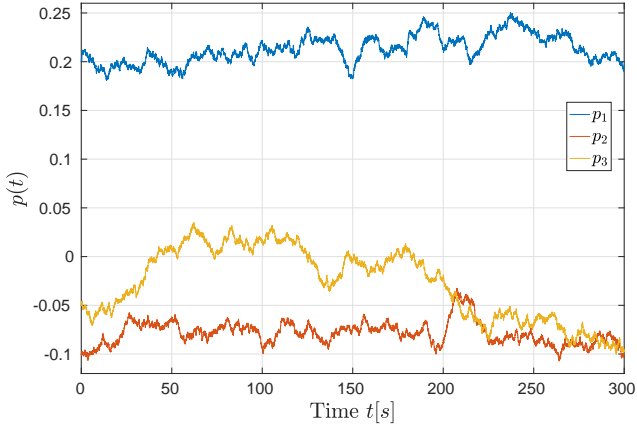}
		\caption{A time-varying torque disturbance 
$\dot{p}=v$, with the initial condition $p(0)=[0.2,-0.1,-0.05]^{T} \ [\mathrm{Nm}]$ and   $v\in \mathbb R^3$, $v_i \in N(0,0.2)$  a  Gauss distribution .}
		\label{fig:P}
	\end{center}
\end{figure}

\begin{figure}[tbh!]
	\begin{center}
		\includegraphics[trim = 10mm 6mm 0mm 0mm,scale=0.34]{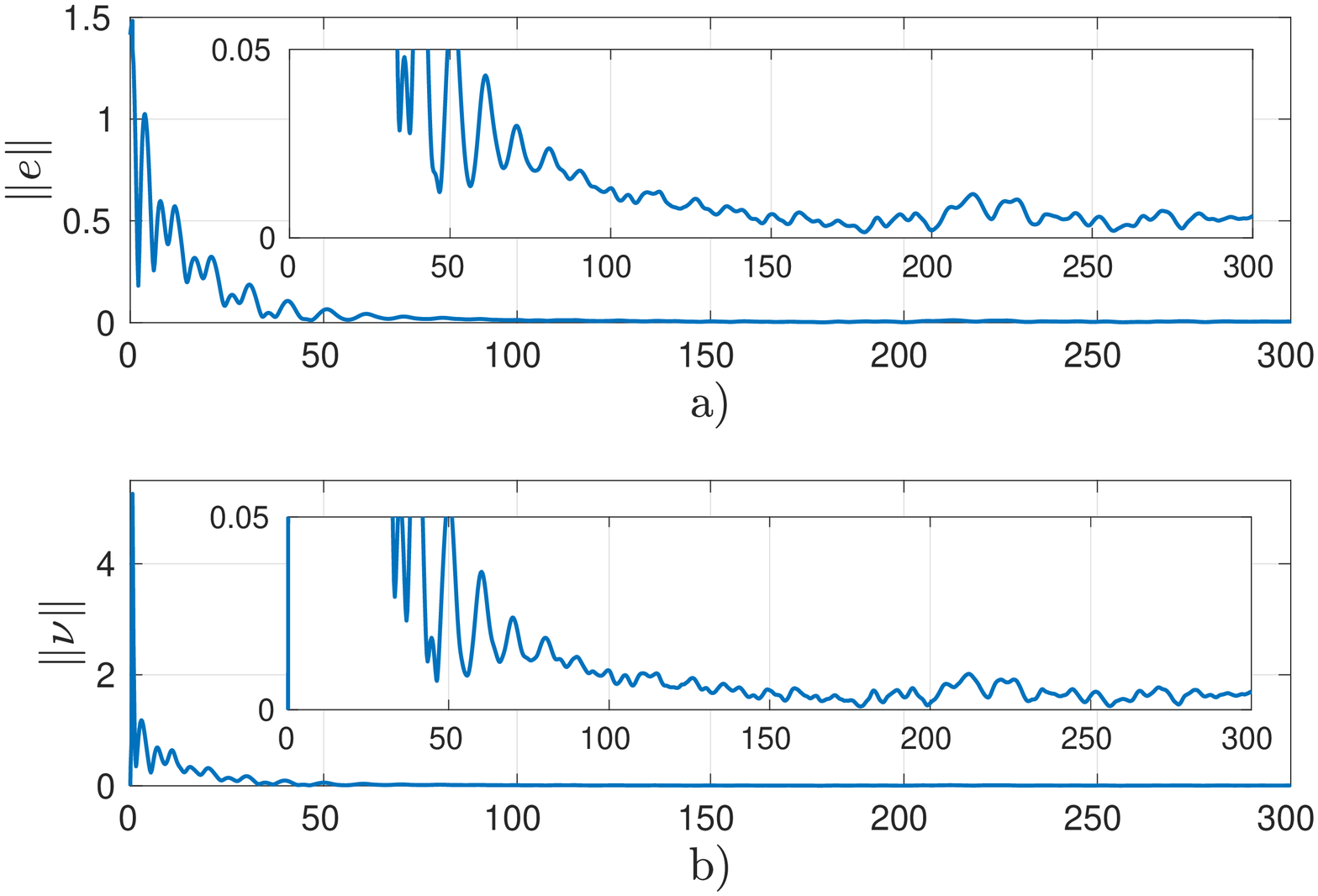}
		\includegraphics[trim = 10mm 6mm 0mm 0mm,scale=0.34]{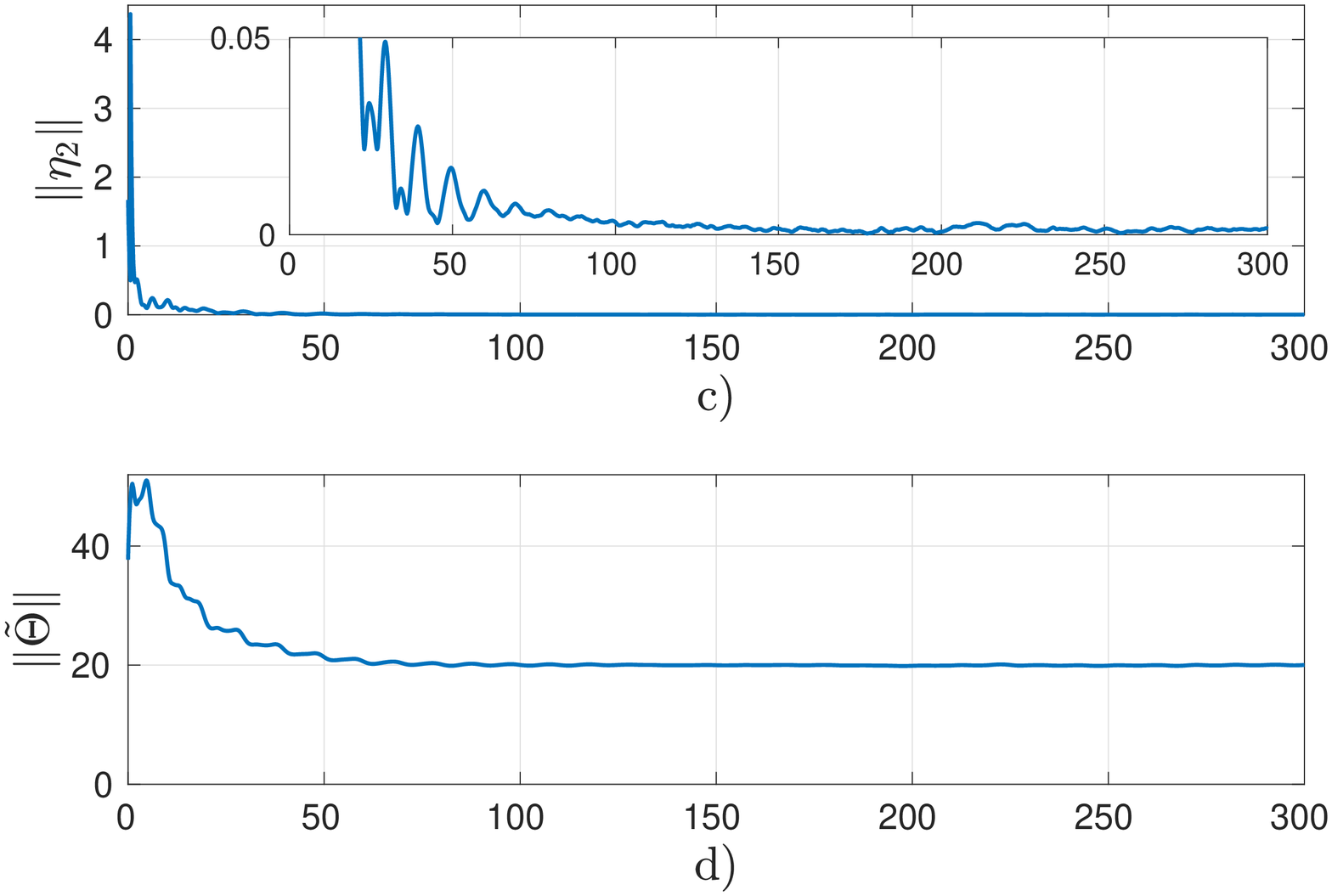}
		\includegraphics[trim = 10mm 6mm 0mm 0mm,scale=0.34]{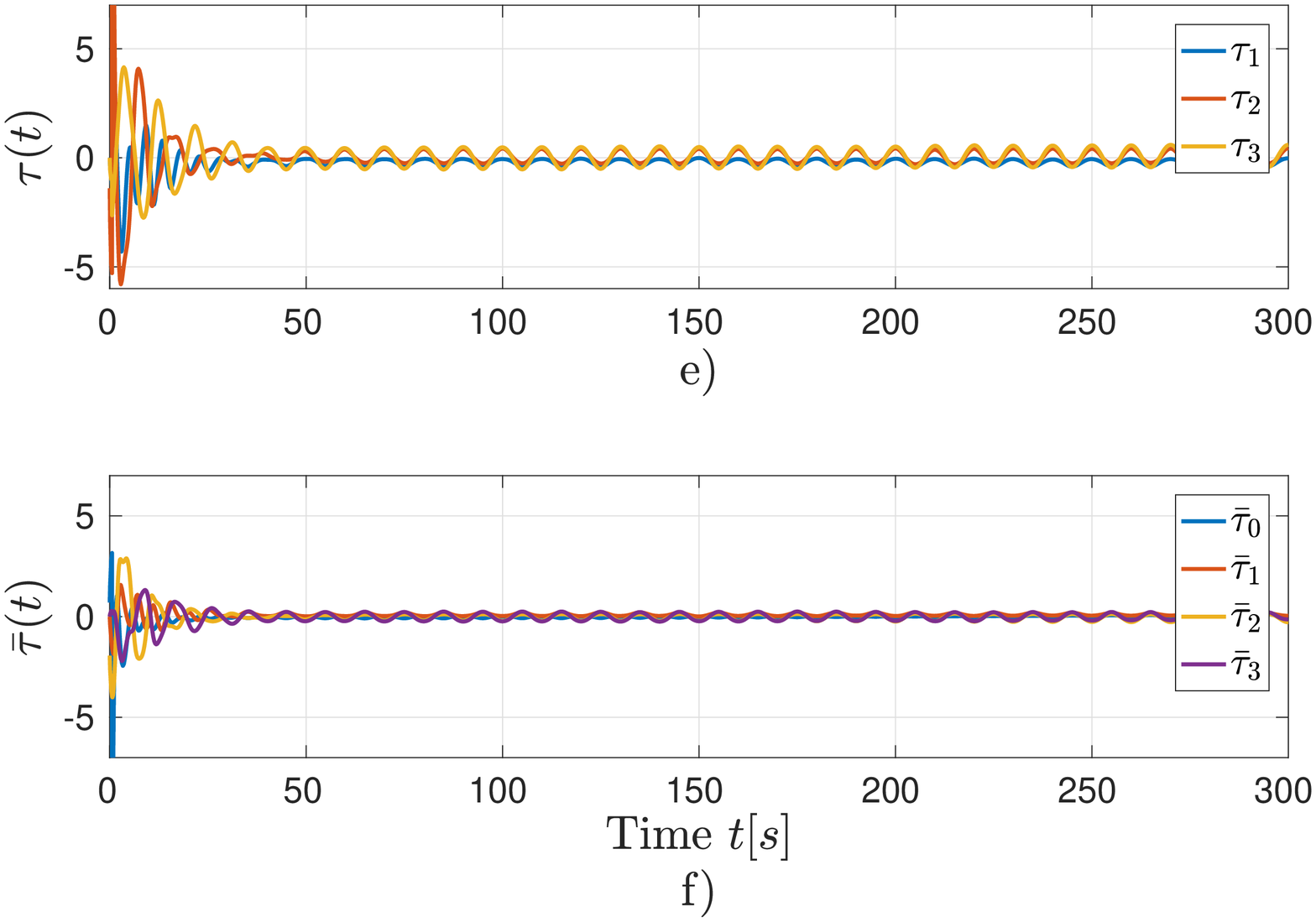}
		\caption{Scenario 2.4: Performance of adaptive controller \eqref{eq:CtrlAd} with gap exceeding $\delta = 0.4$ under time-varying torque disturbance $p(t)$.}
		\label{fig:AdpPv}
	\end{center}
\end{figure}

Although the adaptive hybrid attitude-feedback controller \eqref{eq:CtrlAd} is designed for a constant disturbance, the adaptation law can deal with certain time-varying disturbances as illustrated in Scenario 2.4.  The time-varying disturbance $p(t)$ is given in Table \ref{tab:InitialV} and displayed in Fig. \ref{fig:P}. Likewise, Fig. \ref{fig:AdpPv} shows the performance of the adaptive controller. Notice that the norm of the parameter error $\tilde\Theta (t)$ converges to $20$ as in Scenario 2.3 where a constant disturbance was used . 

\section{Conclusions}\label{Sec:Conc}
This paper has proposed a novel Lagrangian approach to  attitude tracking for rigid spacecraft using unit quaternions. By describing the attitude of a rigid spacecraft  by a 4-DOF Lagrangian dynamics with a holonomic constraint imposed by the norm of a unit quaternion and exploring the energy-conservation property of a Lagrangian system,  this approach enables to leverage a broad class of tracking control designs for mechanical systems based on energy-shaping  methodology to design globally exponentially stable attitude tracking controllers. The salient features of the proposed approach are its capability of using the whole quaternion to design continuous attitude tracking controllers in contrast to using only the vector part of the quaternion reported in the literature, avoiding the topological constraints encountered  in the quaternion group. Since the quaternion representation is singularity-free, this approach also avoids  the singularity issues in any 3-parameter attitude representation like Euler angles, Rodrigues, or Modified Rodrigues parametrizations.

Using this approach, a state-feedback controller was designed when both the attitude and angular velocity are available for feedback. 
Then several important issues, such as robustness to noise in quaternion measurements,  unknown on-orbit torque disturbances,  uncertainty in the inertial matrix, and  lack of angular-velocity measurements are addressed  by designing a hybrid state-feedback controller, an adaptive hybrid state-feedback controller, and an adaptive hybrid attitude-feedback controller. Global asymptotic stability is established for each controller.

%A nonlinear filter to implement the required damping in the state-feedback controller  was used to devise an output feedback controller. Both state feedback and output feedback controllers were  demonstrated to be globally exponentially stable. In the presence of an uncertain inertial matrix, an adaptive attitude feedback controller was also developed, achieving global asymptotic tracking. Numerical simulations were carried out to illustrate the theoretical results and to verify the robustness of the proposed adaptive controller to noisy attitude measurements.

\begin{ack}                               % Place acknowledgements
This work was supported in part by CONACyT under grant 253677 and by PAPIIT-UNAM IN112421, and carried out in the National Laboratory of Automobile  and Aerospace Engineering LN-INGEA.  % here.
\end{ack}

\bibliographystyle{unsrt}
\bibliography{autosam}           % and a bib file to produce the 
%\printbibliography
                                 % bibliography (preferred). The
                                 % correct style is generated by
                                 % Elsevier at the time of printing.

%\begin{thebibliography}{99}     % Otherwise use the  
                                 % thebibliography environment.
                                 % Insert the full references here.
                                 % See a recent issue of Automatica 
                                 % for the style.
%  \bibitem[Heritage, 1992]{Heritage:92}
%     (1992) {\it The American Heritage. 
%     Dictionary of the American Language.}
%     Houghton Mifflin Company.
%  \bibitem[Able, 1956]{Abl:56}
%     B.~C.~Able (1956). Nucleic acid content of macroscope. 
%     {\it Nature 2}, 7--9. 
%  \bibitem[Able {\em et al.}, 1954]{AbTaRu:54}   
%     B.~C. Able, R.~A. Tagg, and M.~Rush (1954).
%     Enzyme-catalyzed cellular transanimations.
%     In A.~F.~Round, editor, 
%     {\it Advances in Enzymology Vol. 2} (125--247). 
%     New York, Academic Press.
%  \bibitem[R.~Keohane, 1958]{Keo:58}
%     R.~Keohane (1958).
%     {\it Power and Interdependence: 
%     World Politics in Transition.}
%     Boston, Little, Brown \& Co.
%  \bibitem[Powers, 1985]{Pow:85}
%     T.~Powers (1985).
%     Is there a way out?
%     {\it Harpers, June 1985}, 35--47.

%\end{thebibliography}

\appendix
\section{Proof of Lemma \ref{lem1}} \label{AppQ}   % Each appendix must have a short title.

Consider the matrix $Q(\cdot)$ defined as \eqref{eq:MatQ}. Let $x\in\mathcal{S}^{3}$, then
\begin{equation*}
\mathrm{det}\left( Q(x) \right) = \left( x^{T}x \right)^{2} = 1.
\end{equation*}
Moreover,
\begin{equation*}
Q^{T}(x)Q(x)=Q(x)Q^{T}(x) = (x^{T}x) I_{4}=I_{4}.
\end{equation*}
This shows the property \ref{pQ1}.

The property \ref{pQ2} can be verified directly by substitution. Indeed, let $x=\left[ x_{0}, \  x^{T}_{v} \right]^{T}$ and $y=\left[ y_0, \ y^T_v \right]^{T}$, with $x_{0}, \ y_0 \in \mathbb{R}$ and $x_{v}, y_{v} \in \mathbb{R}^{3}$. Define $J_{v}(x)= x_{0}I_{3} + S(x_{v}) $ and $J_{v}(y)= y_0 I_{3} + S(y_{v}) $, then
\begin{eqnarray*}
&Q&(y)Q^{T}(x) = 
\left[ 
\begin{array}{cc}
y_0 & -y^{T}_{v} \\ 
y_{v} & J_{v}(y)
\end{array}
\right]\left[ 
\begin{array}{cc}
x_{0} & x^{T}_{v} \\
-x_{v} & J^{T}_{v}(x)
\end{array}
\right] , \\
&=& 
\left[ 
\begin{array}{cc}
y_0 x_{0} + y^{T}_{v}x_{v}  & y_0 x^{T}_{v}-x_{0}y^{T}_{v} + y^{T}_{v}S(x_{v}) \\ 
x_{0}y_{v} - y_{0}x_{v} - S(y_{v})x_{v} & y_{v}x^{T}_{v} + J_{v}(y)J^{T}_{v}(x)
\end{array}
\right]  \\
&=& 
\left[ 
\begin{array}{cc}
y^{T}x  & -z^{T} + y^{T}_{v}S(x_{v}) \\ 
z + S(x_{v})y_{v} & \left( y_{v}x^{T}_{v} - x_{v}y^{T}_{v}\right) +S(z) + \left( y^{T}x \right) I_{3}
\end{array}
\right] ,
\end{eqnarray*}
where $z = x_{0}y_{v}-y_0 x_{v}$ and the property of the skew-symmetric operator $S(u)v = -S(v)u$, $\forall u,v \in\mathbb{R}^{3}$ are  used. Therefore, the matrix $Q(y)Q^{T}(x)$ is skew-symmetric if and only if $y^{T}x = 0$,  which verifies the property \ref{pQ3} and \ref{pQ4}.

Finally, the properties \ref{pQ5} and \ref{pQ6} can be proved straightforwardly as done in \cite{markley2014fundamentals} for the matrix $J(\cdot)$.

\section{Proof of Lemma \ref{lem2}}\label{PfLemma2}         % Sections and subsections are supported  
 The matrix $D(q)$  defined in \eqref{eq:Dq2} is symmetric and positive definite follows straightforwardly because $Q(q)\in SO(4)$  for $q\in\mathcal{S}^{3}$ by  Property \ref{pQ1} of Lemma \ref{lem1}  for  any $m_{0}>0$. This shows the first term  of this  lemma. 

To prove the second term, taking the time derivative of \eqref{eq:Dq2} and substituting \eqref{eq:Cqqp} obtains 
\begin{eqnarray*}
\dot{D}(q) - 2C(q,\dot{q}) &=& Q(\dot{q})M_{0}Q^{T}(q) + Q(q)M_{0}Q^{T}(\dot{q}) \\ 
& & - 2  C(q,\dot{q})  \\
&=& Q(\dot{q})M_{0}Q^{T}(q) + Q(q)M_{0}Q^{T}(\dot{q}) \\
& &+2J(q)S(M\omega )J^{T}(q) \\
& & + 2 D(q)Q(\dot{q})Q^{T}(q). 
\end{eqnarray*}
By the fact that $\dot{q}^{T}q = 0$ and  Property \ref{pQ4} of the matrix $Q(\cdot)$ in  Lemma \ref{lem1}) it follows
\begin{eqnarray*}
\dot{D}(q) - 2C(q,\dot{q}) &=&  Q(\dot{q})M_{0}Q^{T}(q) + Q(q)M_{0}Q^{T}(\dot{q}) \\
& &+2J(q)S(M\omega )J^{T}(q) \\
& & - 2 D(q)Q(q)Q^{T}(\dot{q}) , \\
&=& Q(\dot{q})M_{0}Q^{T}(q) + Q(q)M_{0}Q^{T}(\dot{q}) \\
& &+2J(q)S(M\omega )J^{T}(q) \\
& & + 2 \left( -Q(q)M_{0}Q^{T}(\dot{q}) \right) , \\
&=& Q(\dot{q})M_{0}Q^{T}(q) -Q(q)M_{0}Q^{T}(\dot{q})\\
& &+2J(q)S(M\omega )J^{T}(q),
\end{eqnarray*}
which is skew symmetric because $J(q)S(M\omega )J^{T}(q)$ is skew symmetric. 

\section{Proof of Lemma \ref{lem3}}\label{AppDCp} 
Properties \ref{pLinParam} and \ref{pDC1} are proved in this Appendix, the proof of the rest properties was given    in  \cite{kelly2006control} for robot manipulators and can be shown for the Lagrangian dynamics \eqref{eq:EL-model} following the same procedure as in \cite{kelly2006control} and is therefore omitted here.

{\it Proof of Property \ref{pLinParam}}: For a vector  $u = [ u_{1} ,\; u_{2},\; u_{3} ]^{T}\in \mathbb R^3$ define the map $F\vcentcolon \mathbb{R}^{3}\to \mathbb{R}^{3\times 6}$ as
\begin{equation}\label{eq:Freg}
    F(u) = \left[ \begin{array}{cccccc}
u_{1} & 0 & 0 & 0 & u_{3} & u_{2}  \\
0 & u_{2} & 0 & u_{3} & 0 & u_{1}\\
0 & 0 & u_{3} & u_{2} & u_{1} & 0 
\end{array} \right].
\end{equation}
Then, the product $Mu$ can be expressed as  $Mu = F(u)\theta$,  where $M$ is the inertial matrix in \eqref{eq:InertialM} and  $\theta$ is defined in \eqref{eq:tht}.

Let $w=J^{T}(q)\dot{q}$, then $\dot{w}=J^{T}(q)\ddot{q}$, therefore,  in view of \eqref{eq:Dq2}, it follows that
\begin{align}
    D(q)\ddot{q} &= \left( J(q)MJ^{T}(q) + m_{0}qq^{T} \right) \ddot{q} \notag\\
    &= J(q)F(\dot{w})\theta + m_{0}\left( q^{T}\ddot{q} \right) q. \label{eq:Dqqpp}
\end{align}
In addition,  by using \eqref{eq:Dq2} and Properties \ref{pQ2} and \ref{pQ4} of Lemma \ref{lem1}, the matrix $C(q,\dot{q})$ in \eqref{eq:Cqqp} is rearranged to
\begin{align*}
    C(q,\dot{q}) &= -J(q)S(M\omega)J^{T}(q) - D(q)Q(\dot{q})^{T}Q(q) \\
    &= -J(q)S(M\omega)J^{T}(q) + D(q)Q(q)^{T}Q(\dot{q}) \\
    &= -J(q)S(M\omega)J^{T}(q) \\
    &\quad + D(q)\left( J(q)J^{T}(\dot{q}) + q\dot{q}^{T} \right)  \\
    &= -J(q)S(M\omega)J^{T}(q) \\
    & + \left( J(q)MJ^{T}(q) + m_{0}qq^{T} \right) \left( J(q)J^{T}(\dot{q}) + q\dot{q}^{T} \right).
\end{align*}
By properties \ref{pA1} and \ref{pA3} of the matrix $J(q)$, it follows that
\begin{equation}\label{eq:Cqqp2}
     C(q,\dot{q}) = -J(q)S(M\omega)J^{T}(q) + J(q)MJ^{T}(\dot{q}) + m_{0}q\dot{q}^{T}.
\end{equation}

Then, by \eqref{eq:Cqqp2} and \eqref{eq:AngVel}  the term $C(q,\dot{q})\dot{q}$ can be rewritten  as
\begin{align}
    C(q,\dot{q}) \dot{q} &= \left( -J(q)S(M\omega)J^{T}(q) \right. \notag \\
    &\quad\quad \left. + J(q)MJ^{T}(\dot{q}) + m_{0}q\dot{q}^{T} \right) \dot{q} \notag \\
    &= -J(q)S(M\omega)w + m_{0}\left( \dot{q}^{T}\dot{q} \right) q \notag\\
    &= 2J(q)S(w)Mw +  m_{0}\left( \dot{q}^{T}\dot{q} \right) q \notag \\
    &= 2J(q)S(w)F(w)\theta +  m_{0}\left( \dot{q}^{T}\dot{q} \right) q. \label{eq:Cqqpqp}
\end{align}
Now, by adding \eqref{eq:Dqqpp} and \eqref{eq:Cqqpqp}, it gets
\begin{align*}
    D(q)\ddot{q} + C(q,\dot{q})\dot{q} &= J(q)F(\dot{w})\theta + m_{0}\left( q^{T}\ddot{q} \right) q  \\
    &\quad + 2J(q)S(w)F(w)\theta +  m_{0}\left( \dot{q}^{T}\dot{q} \right) q \\
    &= m_{0} \left( q^{T}\ddot{q} +\dot{q}^{T}\dot{q} \right) q\\
    &\quad + J(q)\left( F(\dot{w}) + 2S(w)F(w) \right) \theta \\
    &= \bar{Y}\left( q,\dot{q},\ddot{q} \right) m_{0} + Y\left( q,\dot{q},\ddot{q} \right) \theta
\end{align*}
which confirms the linear parametrization property \ref{pLinParam} of Lemma \ref{lem3}.

{\it Proof of Property \ref{pDC1}}: The Property \ref{pDC1} can be verified  straightforwardly by taking the time derivative $\dot{D}(q)$ from \eqref{eq:Dq2} and comparing it with $C(q,\dot{q}) + C^{T}(q,\dot{q})$ by using \eqref{eq:Cqqp2}.

\end{document}